\documentclass[apj]{emulateapj}
\usepackage{multirow}

\shorttitle{The First Star Clusters}
\shortauthors{Bland-Hawthorn, Karlsson, Sharma, Krumholz, \& Silk}

\newcommand{\msun}{M_{\odot}}

\newcommand{\tsn}{t_{\rm SN}}
\newcommand{\tcr}{t_{\rm cr}}
\newcommand{\tform}{t_{\rm form}}

\begin{document}

\title{Chemical signatures of the first star clusters in the Universe}

\author{ }
\affil{ }

\author{Joss Bland-Hawthorn\altaffilmark{1}, Torgny Karlsson\altaffilmark{2}, Sanjib Sharma}
\affil{Sydney Institute for Astronomy, School of Physics, University of Sydney, NSW 2006, Australia}
\altaffiltext{1}{Leverhulme Visiting Professor \& Merton College Fellow, University of Oxford, OX1 3RH, UK}
\altaffiltext{2}{Visiting Research Fellow, University of Oxford, OX1 3RH, UK}
\email{jbh@physics.usyd.edu.au}

\author{Mark Krumholz}
\affil{Department of Astronomy \& Astrophysics, University of California, Santa Cruz, CA 95060, USA}

\author{Joe Silk}
\affil{Physics Department, University of Oxford, OX1 3RH, UK}

\begin{abstract}
The chemical abundance patterns of the oldest stars in the Galaxy are expected to contain residual
signatures of the first stars in the early universe. Numerous studies attempt to 
explain the intrinsic abundance scatter observed in some metal-poor populations in terms of
chemical inhomogeneities dispersed throughout the early Galactic medium due to discrete enrichment 
events. Just how the complex data and models are to be interpreted with respect to ``progenitor
yields'' remains an open question. Here we show that stochastic chemical evolution models 
to date have overlooked a crucial fact. Essentially all stars today are born in highly 
homogeneous star clusters and it is likely that this was also true at early times. When this
ingredient is included, the overall scatter in the abundance plane [Fe/H] vs. [X/Fe] (${\cal C-}$space), 
where X is a nucleosynthetic element, can be much less than derived from earlier models. 
Moreover, for moderately flat cluster mass functions ($\gamma \lesssim 2$), and/or for mass functions with
a high mass cut-off ($M_{\rm max} \gtrsim 10^5$ M$_\odot$), stars exhibit a high degree of 
clumping in ${\cal C}-$space that can be identified even in relatively small data samples.
Since stellar abundances can be modified by mass transfer in close binaries, 
clustered signatures are essential for deriving the yields of the
first supernovae. We present a statistical test to determine whether a given set of observations 
exhibit such behaviour. Our initial work focusses on two dimensions in ${\cal C-}$space, but
we show that the clustering signal can be greatly enhanced by additional abundance axes. 
The proposed experiment will be challenging on existing 8-10m telescopes, but relatively
straightforward for a multi-object echelle spectrograph mounted on a 25-40m telescope.
\end{abstract}


\section{Introduction}

The Galactic stellar halo is a vast ancient repository that takes us back to a time when dark matter collapsed into
bound structures and the Galaxy was seeded for the first time. 
Helmi (2008) and Tolstoy, Hill \& Tosi (2009) provide excellent overviews of the many stellar
systems and fragments that inhabit the halo. These include field halo stars (Christlieb et al 2002;
Cayrel et al 2004; Frebel et al 2005; Cohen et al 2007), globular clusters (Gratton, Sneden \& Carretta 2004), dwarf
spheroidals (Mateo 1999; Venn et al 2004), ultra-faint dwarf galaxies (Simon \& Geha 2007; Kirby et al 2008), 
stellar streams (Ibata et al 1995; Chou et al 2010), 
stellar associations (Walsh et al 2007) and satellites to dwarf galaxies (Coleman et al 2004; Belokurov et al 2009).
Already the chemical information arising from the most metal-poor stars is very difficult to unravel
(Nomoto et al 2005; Kirby et al 2008). We have barely begun to understand what these systems are telling us 
about the sequence of events that led to the Galaxy (McWilliam, Simon \& Frebel 2009; Freeman \& Bland-Hawthorn 2002).

The first stars were unique to their time. The first stellar generations changed the universe in many ways, 
in particular, the chemical properties and the
equation of state of the intergalactic medium. But at present there are many unknowns. 
Did the first stars form in isolation or in groups? Were relatively few 
massive stars responsible for reionization or was it triggered by the collective effect of massive star clusters?
Just what are the processes 
that govern star formation at extremely low metallicity? Is this exclusively the domain of the most massive stars, or
can substantial intermediate and low mass stars form (Tsuribe \& Omukai 2006, 2008; Clark, Glover \& Klessen 2008)? 
In other words, did stellar populations observable today exist before reionization (Tumlinson 2010; 
Okrochov \& Tumlinson 2010)?
The first star clusters are of great interest in that they shed light on star formation processes in the early
universe (Bromm, Coppi \& Larson 2002; Abel et al 2002). There are few if any reliable constraints at the present time.

One of the most interesting developments in recent years is the simultaneous measurement of many elemental
abundances for individual metal-poor halo stars or groups of stars (e.g. Beers \& Christlieb 2005). Some of these 
elements, but not all, exhibit intrinsic scatter in the abundance plane [Fe/H] vs. [X/Fe] that exceeds the 
measurement errors. While the scatter is particularly apparent in halo stars (e.g. Roederer et al 2009), 
evidence is now emerging that star-to-star abundance variations exist in dwarf galaxies as well 
(Fulbright, Rich \& Castro 2004; Koch et al 2008; Feltzing et al 2009). The observed scatter is likely to increase now that metal
poor stars are now detected below [Fe/H] $=$ -3 (Norris et al 2009; Starkenburg et al 2010; Simon et al 2010; 
Frebel, Kirby \& Simon 2010).

This has led numerous researchers to argue that the scatter in [X/Fe] is a tracer of an ancient
inhomogeneous medium (e.g. Audouze \& Silk 1995; Ryan, Norris \& Beers 1996; McWilliam 1997; Ishimaru \& Wanajo 1999).
If this interpretation is correct, we would expect that a subset of these stars is telling us something 
fundamental about the first stars and their yields. But not all of the stars are providing us with an 
unambiguous ``reading'' of the early enrichment of the primordial interstellar gas. For example, a significant 
fraction of extremely metal poor stars appear to have undergone mass transfer with a companion (Ryan et al 2005; 
Lucatello et al 2005) which undermines any attempt at inferring the progenitor yields for
elements such as CNO and $\alpha$ elements. Binarity may partly explain why the elemental abundances of
the most metal poor stars defy a clear explanation at the present time (Joggerst et al 2010; McWilliam et al 2009).

We now incorporate a missing ingredient into existing models of stochastic chemical evolution.
In the present-day universe, most stars are born in a single burst within compact clusters and stellar fragments,
rather than in isolation. This fact is well established in the local universe 
(Lada \& Lada 2003) and is likely to be true at the time of the first stars (Clark et al 2008).
Cluster formation has an important consequence for the distribution of stars in the abundance plane.
De Silva et al (2006, 2007a,b) have shown that both old ($\sim$ 10 Gyr) and intermediate-age ($\sim 1$ Gyr)
open clusters are chemically homogeneous to a high degree ($\Delta$[Fe/H] $\lesssim 0.03$ dex). The open cluster
Tombaugh 2 was thought to be a rare example of a chemically {\it inhomogeneous} open cluster (Frinchaboy et al
2008), but this is contradicted by a new study that finds the stellar population to be highly homogeneous 
($\Delta$[Fe/H] $\lesssim 0.02$ dex; Villanova et al 2010). Apart from a few light elements, the same
holds true for globular clusters (Gratton et al 2004), although some systems (e.g. $\omega$Cen) show 
evidence for more than one burst of star formation. Interestingly, even moving groups can show the same 
signature of chemical homogeneity (De Silva et al 2007a; Chou et al 2010; Bubar \& King 2010). 

Bland-Hawthorn, Krumholz \& Freeman (2010) provide a condition that ensures
chemical homogeneity in a young star cluster based on the surface density of the progenitor gas. They
show that chemical homogeneity is expected in open clusters, globular clusters and plausibly clusters
more massive than $10^7$M$_\odot$. {\it This process has not been 
incorporated into stochastic chemical models to date.} Once this is done, we arrive at an important insight
that will enhance the interpretation of metal abundance distributions in individual stellar populations. 
The effects that we highlight can be searched for in existing and in future surveys, as we show.

A key assumption in our present work is that present-day dwarf galaxies are important sites for establishing 
the yields of the first stars and star clusters. This needs some clarification. While it is likely to be true that the most
efficient way to identify metal-poor stars is to target dwarf galaxies, a low value of [Fe/H] is no guarantee
that a star is ancient since it may simply reflect environmental conditions (e.g. low star formation efficiency,
weak gravity field). Published numerical simulations are unclear on whether the most ancient stars are 
solely the preserve of the inner bulge (White \& Springel 2000; Bland-Hawthorn \& Peebles 2006) or spread 
over the entire Galaxy (Scannapieco
et al 2006; Brook et al 2007). But it is now well established that the bulge, the halo and all dwarf galaxies 
comprise stellar populations that are 10 Gyr or older, equivalent to a formation redshift of $z\gtrsim 2$ (e.g.
Tolstoy et al 2009). The relative fractions of dwarf populations that formed before, during or after the
reionization epoch is an open question. We are therefore at liberty to explore new observational constraints
on the nature of the first stars and star clusters.

In \S 2, we present evidence for homogeneous star clusters in the local universe and argue that the same
should be true for their high-redshift counterparts.
In \S 3, we briefly outline the inhomogeneous chemical evolution models that have been developed to date.
In \S 4, we introduce a revised stochastic model of star formation in dwarf galaxies that incorporates the 
`homogeneity' condition 
described by Bland-Hawthorn et al (2010), and show the predictions of the revised model in view of the earlier work.
In \S 5 and in the Appendix, we introduce cluster finding algorithms for both large and small data sets, and demonstrate how the
effects of clustering may already be visible in existing observations. We also explore the longer term prospect offered 
by a multi-object echelle spectrograph on an extremely large telescope (ELT). The conclusions are presented in
\S 6.

\section{Homogeneous star clusters}

Before revisiting stochastic chemical evolution models, we review the key arguments presented in Bland-Hawthorn et al (2010) on the conditions under which star clusters are expected to be highly homogeneous in most elements, as observed for local star clusters (\S 1). We then extend these arguments to clusters at low metallicity.

As a star-forming cloud assembles, turbulent diffusion within it will homogenize its chemical composition \citep{murray90}. For clouds whose turbulent motions are primarily on large scales (as is the case for all local molecular clouds -- \citet{heyer04}), \citet{bland10} show that the time required for this process to smooth out a composition gradient on the size scale of the cloud is roughly  $\tcr = L/\sigma$, where $\tcr$ is the cloud crossing time, $L$ is its size, and $\sigma$ is its velocity dispersion. Smaller-scale gradients are erased more quickly, with the diffusion time varying as the square of the (normalized) characteristic size scale. Since $\tcr$ is comparable to or smaller than the timescale over which star formation takes place, clouds will homogenize as they assemble, and will be pushed away from homogeneity only if supernovae occur during the star formation process, before the cloud disperses. The time required for a very massive star to evolve from formation to explosion defines the supernova time scale, $\tsn \approx 3$ Myr, and we only expect star clusters to be homogeneous if they are assembled on time scales shorter than $\tsn$.

To determine under what conditions this requirement is satisfied, we must compare $\tsn$ to the cluster formation timescale $\tform$. There is considerable debate over this timescale, so for our purposed we will adopt the longest, most conservative proposed timescale of $4$ $\tcr$ (Tan, Krumholz \& McKee 2006). Since
\begin{equation}
\tcr = \frac{0.95}{\sqrt{\alpha_{\rm vir} G}}\left(\frac{M}{\Sigma^3}\right)^{1/4}
\end{equation}
for a cloud of mass $M$, column density $\Sigma$, and virial ratio $\alpha_{\rm vir}$; observed clouds have $\alpha_{\rm vir}\approx 1.5$ \citep{mckee07b}. For convenience we write the final stellar mass of a cluster as $M_* = \epsilon M$, where $\epsilon$ is the star formation efficiency. Both observational and theoretical arguments suggest $\epsilon \approx 0.2$ independent of $M$ (Lada \& Lada 2003; Fall, Krumholz \& Matzner 2010). Thus the condition that $\tform \approx 4 \tcr < 2\tsn$ (where the factor of 2 arises because the typical star forms halfway through the formation process) is satisfied only if
\begin{equation}
M_{*,5}^{1/4} \Sigma_0^{-3/4} < 2.8,
\end{equation}
where $M_{*,5} = M_*/(10^5\,M_\odot)$ and $\Sigma_0 = \Sigma/(1\mbox{ g cm}^{-2})$, and we have adopted fiducial values of $\epsilon=0.2$ and $\alpha_{\rm vir} = 1.5$. Star-forming regions within the galaxy have $\Sigma_0 \approx 1$ g cm$^{-2}$ independent of mass (Fall et al.\ 2010). Globular clusters today have somewhat higher values of $\Sigma$, although it is unclear if this reflects the conditions under which they formed, or is the result of dynamical evolution since their formation. Regardless, this analysis suggests that clusters with masses up to $\sim 10^5$ $\msun$ should be chemically uniform and maybe much higher if systems form with the nuclear densities observed in today's globular clusters (Bland-Hawthorn et al 2010). 

There are few observational constraints on the existence of star clusters in extremely metal poor gas.
In the nearby universe, clear evidence for star clusters at [Fe/H]$\;\approx\;$-1.7 is observed in the most metal-poor, blue compact
dwarf galaxy I Zw 18 (Izotov \& Thuan 2004). Globular clusters are known to exist down to
[Fe/H]$\;\approx\;$-2.4 (Gratton et al 2004). We now argue that the above estimate for the uniform mass limit is likely to apply at metallicities as low as $[{\rm Fe}/{\rm H}] \approx\;$-5. 

Low metallicity can change the star formation process in two ways that are relevant to the chemical signatures of the resulting stars. First, a change in metallicity can affect the way star-forming clouds fragment; if there is no fragmentation down to sub-solar masses below a certain metallicity, then no stars below that metallicity will survive to the present day. Fragmentation of low metallicity gas has received extensive attention in the literature, which we will only summarize here. The main point relevant for our purposes is that, when dust cooling is considered, gas is able to fragment to sub-solar masses even at metallicities at low as $\sim 10^{-6}$ of the solar abundance (Clark et al.\ 2008; Schneider \& Omukai 2010). Once fragmentation is possible, turbulence naturally generates a mass spectrum of fragments with a slope comparable to the Salpeter slope $dn/dm \propto m^{-2.35}$ (Padoan \& Nordlund 2002; Hennebelle \& Chabrier 2008), and the properties of the turbulence do not depend on the redshift or the metallicity. Alternately, Clark et al.\ (2009) have proposed that competitive accretion processes would generate a universal mass spectrum, although doubts have been raised about whether this process in fact operates on both theoretical (Krumholz, McKee \& Klein 2005) and observational (Andr\'e et al.\ 2007) grounds. In either case, however, we expect there to be some sub-solar mass stars formed that can survive to the present day.

Given that small stars form at low metallicity, we can then ask whether our conclusions about cluster chemical homogeneity will continue to apply in this regime. A failure of homogeneity could occur either if clouds did not homogenize during star cluster formation, or if the cloud properties changed such that the cluster formation time became longer than the supernova timescale. The former is unlikely because the homogenization time is comparable to the crossing time, and it is implausible that any star formation process could take place in less time. The latter would require that protocluster gas clouds all have surface densities significantly below $\sim 0.1$ g cm$^{-2}$ which is highly unlikely to be true. However this would imply that the star-forming clouds had surface densities at or below the {\it mean} surface density of observed high redshift galaxies (e.g.\ Genzel et al.\ 2006), which is implausible. We therefore conclude that homogeneity should continue to apply as well.\footnote{Homogeneity could also fail in very massive clusters, those larger than $\sim 10^7$ $\msun$ (Maraston et al 2004; Larsen 2009). However in this case even if the whole clusters were not chemically homogeneous, smaller mass portions within them would be, and there is little distinction from the standpoint of the model we present below. The effect would simply be to break up very large clusters into a number of chemically homogeneous small clusters. Given that large clusters are rare, this would not affect our signal significantly.}

\section{Existing inhomogeneous chemical evolution models}

\subsection{A statistical treatment}

Many authors have attempted to explain the declining scatter in elemental abundances with increasing
[Fe/H] in terms of star formation within an ISM that is enriched by a succession of SN events
(Ishimaru \& Wanajo 1999; Shigeyama \& Tsujimoto 1998; Raiteri et al 1999; Argast et al 2000, 2004; 
Wasserburg \& Qian 2000; Karlsson \& Gustafsson 2001, 2005;
Qian 2000, 2001; Travaglio, Galli \& Burkert 2001; Fields, Truran \& Cowan 2002). In the [Fe/H] vs. [X/Fe] 
plane,\footnote{In the standard notation, for a given star with a measured abundance ratio $({\rm X/Fe})_*$,
it is convenient to write [X/Fe] = $\log_{10}({\rm X/Fe})_* - \log_{10}({\rm X/Fe})_\odot$ which is a logarithmic scale
normalized to solar abundances. } the scatter in [X/Fe] is due to two or more classes of SN that
produce distinct yields of [X/Fe]. In these stochastic models, the scatter converges roughly quadratically 
to a mean value of [X/Fe] given by the supernova yields weighted by the initial mass function (IMF).
In its simplest form, this behaviour is driven by the number of SN events ($n_{\rm SN}$), such that 
[Fe/H] = [Fe/H]$_{\rm min} + \log n_{\rm SN}$, where [Fe/H]$_{\rm min}$
is the minimum allowed metallicity. In other words, if [Fe/H]$_{\rm min} = -5$ consistent with the most metal-poor
stars to date (e.g. Frebel et al 2010), it takes 10$^5$ supernovae to enrich a gas parcel to solar abundance.
An example of this behaviour is presented in Bland-Hawthorn \& Freeman (2004; Fig. 3) where
the number of SNe are indicated.

If the interpretation of declining inhomogeneity is broadly correct, we can hope to learn about the yields of the first
SNe responsible for the enrichment (Fields, Truran \& Cowan 2002). In other words, if the scatter in [X/Fe] is due to a high-yield source (class A) and
a low-yield source (class B), an individual star with known [Fe/H] has a fraction of $f_A = n_A/n_{\rm SN}$ 
progenitors and $f_B= n_B/n_{\rm SN}$ progenitors of a total progenitor population of 
$n_{\rm SN} = n_A + n_B$ supernovae.
We can simulate this with a random variable $f=f_A$ drawn from the beta distribution function,
\begin{equation}
\label{eqn:beta}
\xi(f;\alpha,\beta) = {{\Gamma(\alpha+\beta)}\over{\Gamma(\alpha)\Gamma(\beta)}}f^{\alpha-1}(1-f)^{\beta-1}
\end{equation}
defined for $f \in [0,1]$ such that $\int_0^1 \xi\: df = 1$ and 
where $\Gamma$ is the gamma function. The quantities $\alpha$ and $\beta$ describe 
the yields of the two populations, such that
$\alpha=\bar{f_A}(n_{\rm SN} -1)$ and $\beta = (1-\bar{f_A})(n_{\rm SN} -1)$. 
The mean, variance and higher moments of the 
normalised distribution $\xi(f)$ depend only on $\alpha$ and $\beta$. 

Equation~\ref{eqn:beta} has several remarkable properties that are useful for describing the declining influence
of chemical inhomogeneities in the early ISM. The mean is given by $\mu = \alpha(\alpha+\beta)^{-1} = \bar{f_A}$
such that $\bar{f_A}$ is the mean of the distribution $\xi(f_A)$. 
The variance is given by $\sigma^2 = \alpha\beta(\alpha+\beta)^{-2}(\alpha+\beta+1)^{-1}$
which leads to a scatter that declines as $\sigma \propto n_{\rm SN}^{-0.5}$ as expected\footnote{Note that
in the abundance plane, $\sigma({\rm [Fe/H]}) \propto 10^{-0.5\rm [Fe/H]}$.}.
Even initially skewed distributions in $\xi$ converge to the normal distribution ($\mu = \bar{f_A}$)
in the limit of high $n_{\rm SN}$, as expected from the central limit theorem.

\begin{figure}
\plotone{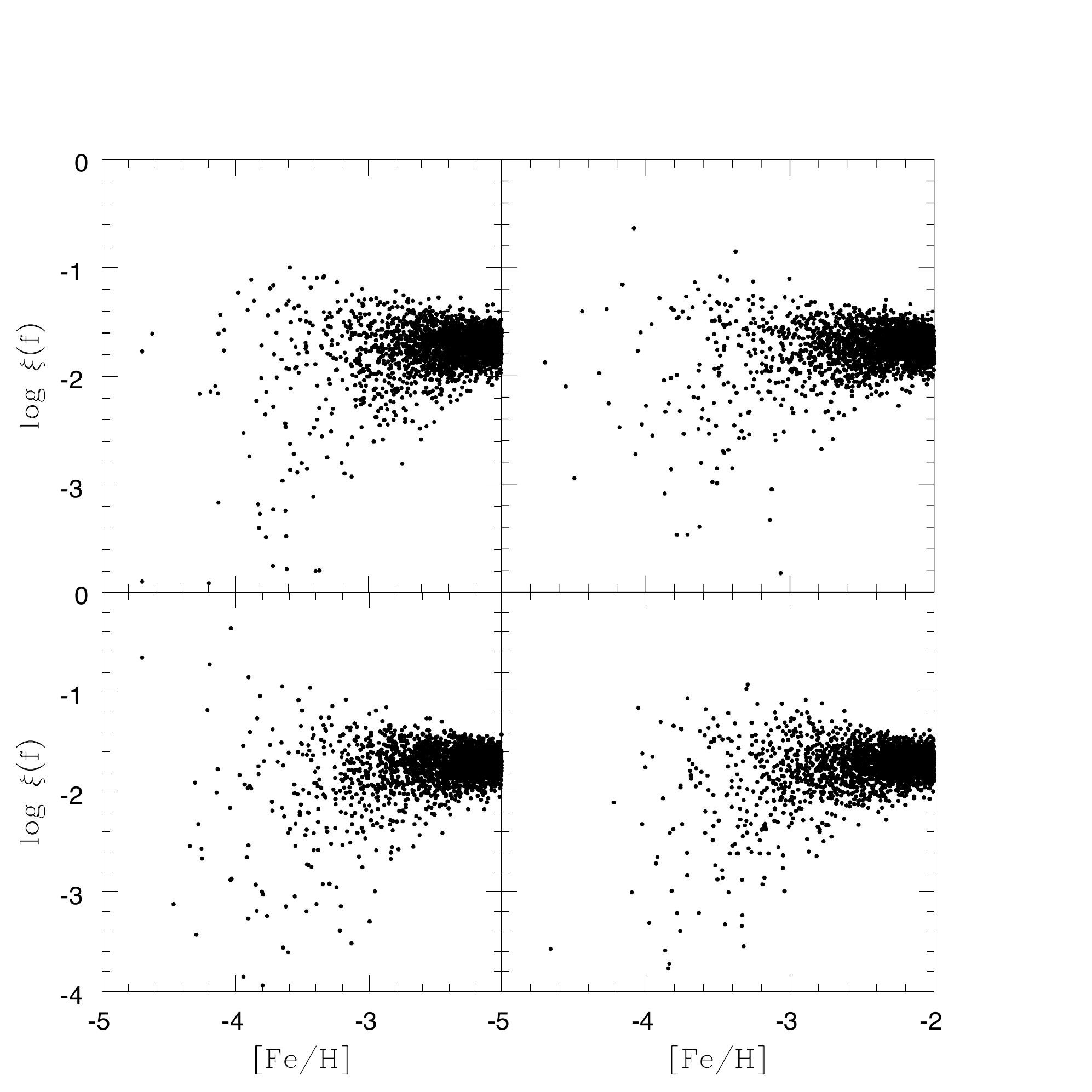}
\caption{\label{beta}
Four realizations of the beta distribution in equation~\ref{eqn:beta}. The vertical axis $\log\xi$ can be trivially rescaled
to match the observed scatter in [X/Fe]. The abundance scatter declines quadratically (in linear space) as expected (see \S3.1).
}
\end{figure}

Four realizations of the beta distribution (i.e. no clustering) are presented in
Fig.~\ref{beta}; a total of $n_* = 3000$ points was used for each simulation distributed in [Fe/H] according to 
Fig.~\ref{MDF} (see \S 5).
The $\log\xi$ distribution can be renormalized trivially to match a given set of [Fe/H] vs. 
[X/Fe] observations (e.g. Fields et al 2002). 
The beta distribution $\xi$ is everywhere continuous in $f_A$, and therefore $f_B$, 
except in the limit of small $n_A$ or $n_B$. This recognizes the fact that,
as stated by Fields et al (2002), ``a given parcel of ISM
gas and dust can be enriched to different degrees by the ancestors it had." 
But in the limit of small $n_A$ or $n_B$, discreteness effects may exist.
Stars with very few prior enrichments, if they can be identified, provide crucial information on the progenitor yields of class A or B
sources (e.g. Shigeyama \& Tsujimoto 1998; Karlsson \& Gustafsson 2001; Ballero et al 2006),
but within the domain of the statistical model, these are rare events 
and may be difficult to find unambiguously. This is an issue we return to in \S 5.
At high [Fe/H], when new sources of metals become dominant (e.g. type Ia SNe,
asymptotic giant branch stars [AGB]), the simple description in equation~\ref{eqn:beta} breaks down.

To recap, the model described by equation~\ref{eqn:beta} assumes that a star with a given 
[Fe/H] has been enriched to different degrees by its type A and type B ancestors which is considered to be a
continuous rather than a discrete process. Equation~\ref{eqn:beta} is used to generate a distribution of possible values of $f_A$ at a 
constant value of [Fe/H] or equivalently $n_{\rm SN}$. More complex models show that there is generally no simple relation 
between $f_A$ and [Fe/H] (e.g. Qian 2001; Bland-Hawthorn \& Freeman 2004).

We note already from Fig.~\ref{beta} that
spurious groupings (false positives) are inevitable, particularly when the data points are convolved 
with typical errors of 0.1 dex psf for differential abundance analysis. The importance of the beta distribution in equation~\ref{eqn:beta} 
is that it enables us to efficiently generate large numbers of unbiassed realizations ($\gtrsim 10^3$) of the theoretical abundance plane,
something that is infeasible with full-blown chemical modelling. These are required to
calibrate the statistical properties of our group-finding algorithm in \S 5. This step is necessary if we are to identify
significant groups against a rapidly changing background, as observed in Fig.~\ref{beta}. Clusters at low [Fe/H] will have
a higher level of significance than abundance groupings at higher [Fe/H] (cf. Karlsson et al 2008) and we need to 
be able to identify groups with smaller number statistics.

But first we must consider more complex treatments of stochastic chemical evolution that will allow us
to include the effects of clustering during star formation. 

\subsection{Inhomogeneous stochastic model}

The chemical evolution model that most resembles our new work is Argast et al (2000, 2004). Their model was computed
over a volume (2.5 kpc)$^3$ for a uniform gas mass of 10$^8$M$_\odot$ at a resolution of (50 pc)$^3$, such that
they were sensitive to chemical inhomogeneities on mass scales as low as 10$^3$M$_\odot$. A distinct feature of
the Argast model is that star formation mostly occurs in the expanding shell of material swept up by the supernova
shock front. They make the simplifying assumption that [X/Fe] is determined by the SN yield such that
[Fe/H] is a combination of the SN yield and the swept-up ISM, and that the shell is everywhere fully mixed.

There are several problems with this assumption.
First, as far as we are aware, there is no compelling evidence for star formation occurring in supernova remnant
shells. The closest one gets are some studies of the Large Magellanic Cloud and the Galaxy
(Yamaguchi et al 1999, 2001) which conclude that maybe $\sim$10\% of star forming regions today appear to
be triggered by local supernova events. There is a clear distinction that must be made. If the star formation occurred 
in collapsing massive clouds triggered by the
passing shock wave, rather than in the swept up shell, we would still expect these star clusters to be highly homogeneous.

Secondly, core-collapse supernovae are inherently asymmetric due to the nature of the 
explosion mechanism due to stellar rotation, magnetic fields, and so on (Wang \& Wheeler 2008). 
Asymmetric ejections from these supernovae are well supported by observations (q.v. Maund et al 2009).
Therefore, the remnant shells are highly unlikely to be chemically homogeneous, again supported
by x-ray observations of nearby supernova remnants (Wang \& Wheeler 2008). Asymmetric enrichment
would have led to even larger spread in the elemental abundances, thereby invalidating their
model comparisons.

Thirdly, our new homogeneity condition, supported by observation, is essentially independent of the original gas distribution. This constitutes a smoothing scale in mass below which metallicity variations cannot occur, thereby suppressing the
amount of scatter observed in the abundance plane (cf. Argast et al 2004).

We now introduce our revised chemical evolution model which incorporates the onset of chemically 
homogeneous star clusters.

\section{Revised chemical evolution model}

\subsection{Initial cluster mass function}

In order to derive the impact of star clusters on the abundance plane, we must consider the
progenitor mass distribution of star clusters.
It is now well established that star clusters have a range of masses that extend from a maximum
mass ($M_{\rm max}$) to a minimum mass ($M_{\rm min}$). The form of the birth distribution
is known as the initial cluster mass function (ICMF) and is assumed to have the form  
\begin{equation}
\mathrm{d}N/\mathrm{d}M_{*} = \chi(M_{*}) = \chi_0 M_{*}^{-\gamma}.
\label{icmf}
\end{equation}
The observations may support a universal slope of $\gamma \approx 2$ in all environments, i.e. 
equal mass per logarithmic bin (Fall, Chandra \& Whitmore 2005, 2009; Lada \& Lada 2003; Elmegreen 2010). 
In this picture, the only parameter that does appear to vary at all between galaxies today is $M_{\rm max}$, which can be 
represented schematically as a Schechter function-like cutoff in the ICMF, although there is no good
reason to prefer this functional form over a simple truncation.

A power law is logarithmically divergent at both low and high masses, so it must be truncated somewhere. At low masses, the truncation is due to the discreteness of stellar masses -- the smallest cluster is simply one star. At large masses, there must also be a truncation. For this reason, and because the molecular cloud mass function (as opposed to the cluster mass function) is observed to have a non-trivial truncation, there is likely to be a maximum cluster mass, and that it varies depending on galactic environment. In disk galaxies, a possible explanation
is that the truncation mass is of order the Toomre mass in the galactic disk (Toomre 1964; Escala \& Larson 2008), which is 
$\sim 10^6$M$_\odot$ for the Milky Way, but is significantly larger for present-day starburst/merger galaxies 
(Larsen 2009) and their high-redshift counterparts (Genzel et al 2006; Forster-Schreiber et al 2009). The situation in spheroidal
galaxies is much less clear but the common occurrence of globular clusters indicates that massive cluster formation must
take place (Elmegreen 2010).

\subsection{Stochastic chemical evolution}
\label{model}

We take as our starting point the stochastic chemical enrichment model presented in Karlsson (2005, 2006) and Karlsson et al. (2008). 
This model is now updated to include the homogenizing processes that must occur during the formation of star clusters (\S 2). But in order to do this,
two additional mixing processes must be accounted for. First, since stars within clusters show no evidence of scatter in chemical abundance ratios, the gas involved in the formation of a cluster must be homogenized prior to, and stay homogenized during, the formation of the individual stars. We include a 
mixing process to ensure that this is always true during the cloud collapse and cluster formation phase.
Second, since massive stars are short lived ($\tau_{\star} \lesssim 20$ Myr) they will explode as supernovae (SNe) before the cluster has dispersed. Assuming that the cluster is unbound and that the intrinsic velocity dispersion of the stars in the cluster is $\sim 1$ km s$^{-1}$, massive stars will, on average, be $\sim 20$ pc apart when they go off as SNe. This is less than the typical size ($\sim 100$ pc) of a SN remnant (SNR) as it merges with the ambient medium. (e.g. Ryan et al. 1996). Therefore, we assume that the ejecta of all SNe formed within a single cluster
enrich the ISM collectively with a mixing mass that scales linearly with the total energy output of the clustered SNe.
These two processes, which produce additional averaging of newly synthesized material before a new generation of stars is formed, have not been considered in earlier models.

The mixing volume, $V_{\mathrm{mix}}^{\mathrm{e}}$, of each chemical enrichment event is given as a power-law expression. 
The present model is somewhat simplified as we suppress the continuous mixing due to the bulk random motions in the turbulent ISM.
Without the time-dependent turbulent diffusion term (cf. Karlsson et al. 2008, equation 1), $V_{\mathrm{mix}}$  reduces to
\begin{equation}
V_{\mathrm{mix}}^{\mathrm{e}}(k) = \frac{4 \pi}{3}\sigma_{\mathrm{E}}(k)^{3/2} = M_{\mathrm{dil}}(k)/\rho,
\label{vmixce}
\end{equation}
where 
\begin{equation}
\sigma_{\mathrm{E}}(k) = \left(\frac{3M_{\mathrm{dil}}(k)}{4 \pi \rho}\right)^{2/3},
\label{sigmae}
\end{equation}
$\rho$ is the density of the ISM and $M_{\mathrm{dil}}(k)$ is the dilution mass of the ejecta of a number $k$ ($\equiv n_{\rm SN}$) of SNe
as they merge with the ambient medium (e.g. Cioffi, McKee \& Bertschinger 1988). Here, we assume that the total dilution mass for the ejecta of multiple SNe exploding in a cluster is proportional, on average, to the total energy released by the SNe associated with that cluster, such that 
\begin{equation}
M_{\mathrm{dil}}(k) = \sum\limits_{j=0}^{k}M_{\mathrm{sw}}^j,
\label{Mdil}
\end{equation}
where $M_{\mathrm{sw}}^j$ is the mass swept-up by individual SNe assuming it explodes in isolation. In order to introduce a small amount of randomness, $\log(M_{\rm sw})$ is drawn from a normal distribution centred on $\log(M_{\mathrm{sw}}) = 5$ with a width of $0.25$. Note that this mixing process is denoted a single enrichment event, even though it may involve enrichment by multiple SNe.  

In addition to this mixing, there is also a mixing volume associated with the formation of the cluster during the collapse of the molecular cloud, $V_{\mathrm{mix}}^{\mathrm{f}}$, such that
\begin{equation}
V_{\mathrm{mix}}^{\mathrm{f}} = M/\rho,
\label{vmixcf}
\end{equation}
where $M$ is the mass of the molecular cloud associated with the star cluster. Within this volume, everything is assumed to be thoroughly mixed before stars are allowed to form. 

In our models, we explore the range $1 \lesssim \gamma \lesssim 2.5$ (e.g. Kroupa \& Boily 2002), between clusters of mass $(M_{\rm min}, M_{\rm max}) = (5\; {\rm M}_\odot, 5\times10^4\; {\rm M}_\odot$), which is our fiducial mass range (Larsen 2009; Portegies Zwart et al 2010).
We adopt a simple scaling relation between the cluster mass and the mass of the parent molecular cloud, such that $M = M_{*}/\epsilon$, where we adopt a star formation efficiency of $\epsilon = 0.2$ (see \S 2). Similarly, the stellar initial mass function (IMF) is governed by 
\begin{equation}
\mathrm{d}n/\mathrm{d}m = \phi(m) = \phi_0m^{-\alpha}.
\label{imf}
\end{equation}
For simplicity, $\alpha = 2.35$, which recovers the Salpeter IMF. The range of stellar masses is set to $0.1\le m/{\rm M_{\odot}} \le 100$ and $\phi_0 = 6.03\times10^{-2}~{\rm M_{\odot}}^{1.35}$.

Formally, the average number of clusters contributing to the chemical enrichment in a random point in space can be expressed by the parameter $\mu_{\mathrm{e}}$, here given by
\begin{equation}
\mu_{\mathrm{e}}(t) = \int\limits_0^t \sum\limits_{k=0}^{k_{\mathrm{max}}} a_k\times V_{\mathrm{mix}}^{\mathrm{e}}(k) u_{\mathrm{cl}}(t')\mathrm{d}t'
\label{muce}
\end{equation}
where  $a_k = N_k/\sum_{k=0}^{k_{\mathrm{max}}}N_k$ is the fraction of clusters in which $k$ SNe explode, while $u_{\mathrm{cl}}(t')$ is the formation rate of star clusters,  closely related to the star formation rate and $V_{\mathrm{mix}}^{\mathrm{e}}(k\!\!\!=\!\!\!0)=0$. The value of $k_{\mathrm{max}}$ is set by the IMF and the upper mass limit of star clusters. On average, $270$ massive stars explode as SNe in a cluster of mass $5\times 10^4~{\rm M_{\odot}}$. Such a cluster will not form more than $320$ SNe at the $3\sigma$-level. We set $k_{\mathrm{max}}=400$ to be on the safe side.

Now, if we make the simplifying assumtion that star clusters are randomly distributed in space (i.e., not themselves clustered), the probability of finding a region in space enriched by $\kappa$ events, i.e., $\kappa$ clusters producing one or more SNe, at time $t$ is given by the Poisson distribution
\begin{equation}
P(\kappa,\mu_{\mathrm{e}}(t))=e^{-\mu_{\mathrm{e}}(t)}\mu_{\mathrm{e}}(t)^{\kappa}/\kappa!
\label{poisson}
\end{equation}

To follow the chemical enrichment in a typical dwarf galaxy, we assume a simulation box of $1.6$ kpc on a side, with an initial, constant particle density of $n_0=1$ cm$^{-3}$, corresponding to a gas density of $\rho_0 = 2.06\times 10^{-24}$ g cm$^{-3}$. The initial mass of baryons in the box is thus $3.1\times10^7~{\rm M_{\odot}}$ which is sufficient to make the largest star clusters considered in this work. The initial metallicity is set to $Z=0$.  In the box, star clusters are allowed to form from collapsing molecular clouds. The molecular clouds are distributed randomly within the box and the masses of the corresponding clusters are distributed according to equation (\ref{icmf}). During the collapse of a molecular cloud, the volume, $V_{\mathrm{mix}}^{\mathrm{f}}$, of gas corresponding to the mass of the cloud is made chemically homogeneous.  The stars of the cluster will all have chemical abundances equal to those of the parent cloud. The number of massive stars, $k$, in the cluster that will explode as SNe is, again, determined by the Poisson statistics $P(k,\mu_{\mathrm{SN}})$, where the mean number of SNe in a cluster of a given mass is given by
\begin{equation}
\mu_{\mathrm{SN}} = \epsilon M  f_{\mathrm{SN}}/\overline{m},
\label{musn}
\end{equation}
where $\epsilon$ is the star formation efficiency, $M$ is the mass of the molecular cloud, $f_{\mathrm{SN}}=1.9\times10^{-3}$ is the fraction of massive stars exploding as SNe, and $\overline{m}=0.35~{\rm M_{\odot}}$ is the mean stellar mass in a single stellar population. The last two parameters are determined by the IMF. For large values of $\mu_{\mathrm{SN}}$, the Poisson distribution approaches a normal distribution.  We make use of this fact to simplify the calculations for $\mu_{\mathrm{SN}} > 32$. In clusters where one or more SNe are formed, the ejecta of all SNe, with masses distributed according to equation~\ref{imf}, are homogeneously mixed with the metals already present within the volume $V_{\mathrm{mix}}^{\mathrm{e}}$.  The Fe-core collapse SN yields, in particular of Ca and Fe, are taken from Nomoto et al. (2006) while the yields of Eu, representing r-process elements and assumed to be formed in O-Ne-Mg core collapse SNe in the mass range $8\le m/{\rm M_{\odot}}\le 10$, are taken from Argast et al. (2004). In order to match the simulations with observations of metal-poor stars in the Galactic halo, the yields of Eu are doubled.

In order to account for the amount of mass locked up in low-mass stars and stellar remnants and the mass lost due to star formation driven galactic outflows, a fraction, $f_{\mathrm{out}}=0.46$, of the mass of each star-forming molecular cloud is subtracted from the total mass of the system.  The gas density is decreased accordingly. Infall and mass lost via interaction with the surroundings, such as tidal and ram-pressure stripping are not considered here. When the total mass and gas density of the system has been re-calculated, the next cycle begins with the formation of a new cluster. Since our focus is in the metal-poor regime, the yields of type Ia SNe and AGB stars are not considered here. 

\subsection{Results}

\begin{figure}
\plotone{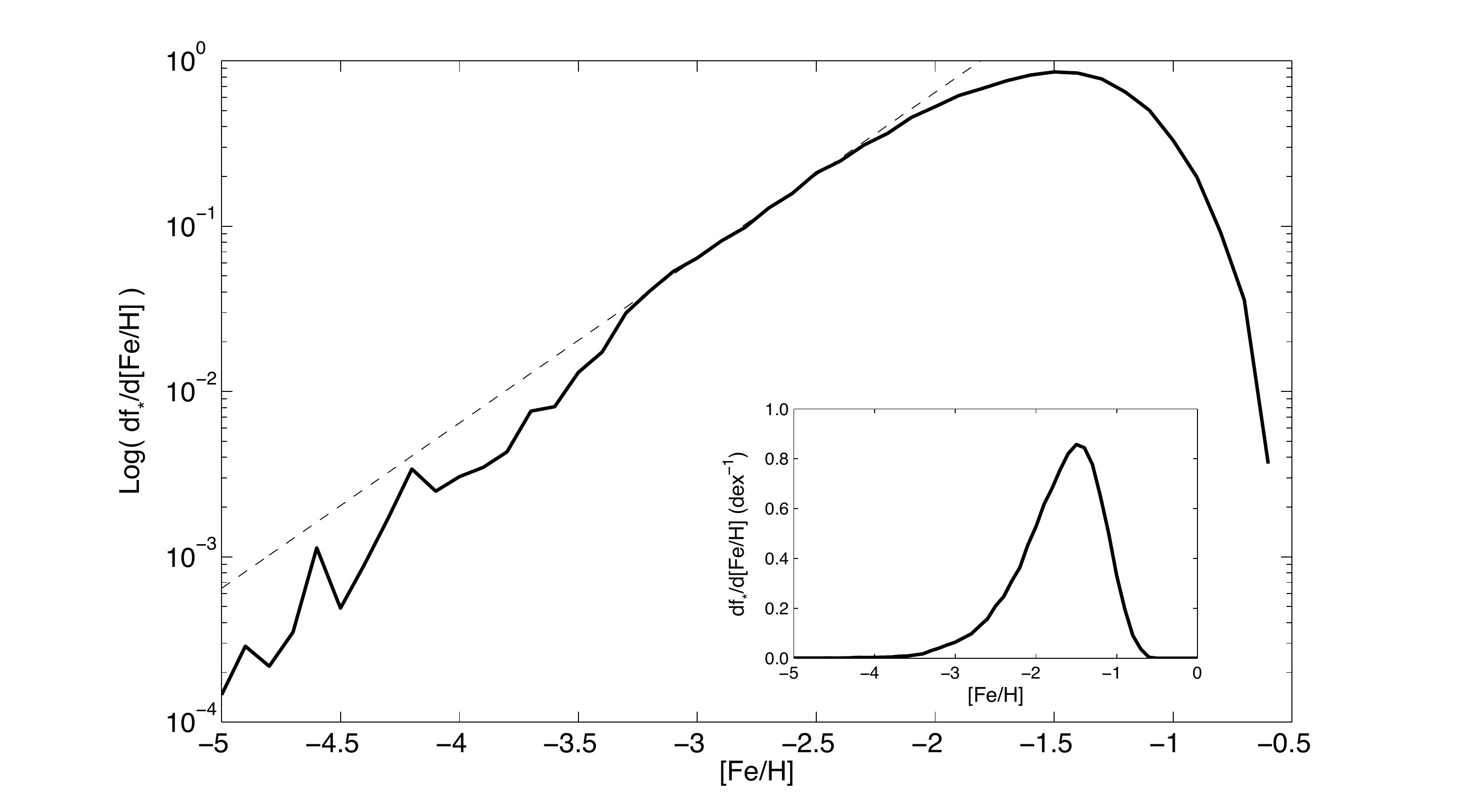}
\caption{ \label{MDF}
Metallicity distribution function (MDF) for the stochastic chemical evolution models presented in Figs.~\ref{CaFeH}
and \ref{rFeH}. The main plot shows the log of the MDF and the inset is the more conventional linear MDF.
The quantity $f_*$ is the fraction of stars that fall within each [Fe/H] bin (1 dex).
The dashed curve shows the gradient defined by $\log {\rm d} f_* / {\rm d[Fe/H]} = 1$ which illustrates that
the fraction of stars in each [Fe/H] bin increases roughly by a factor of 10 as the metallicity increases up to the
turnover at $[Fe/H] \approx -1.5$.
}
\end{figure}

\begin{figure}
\plottwo{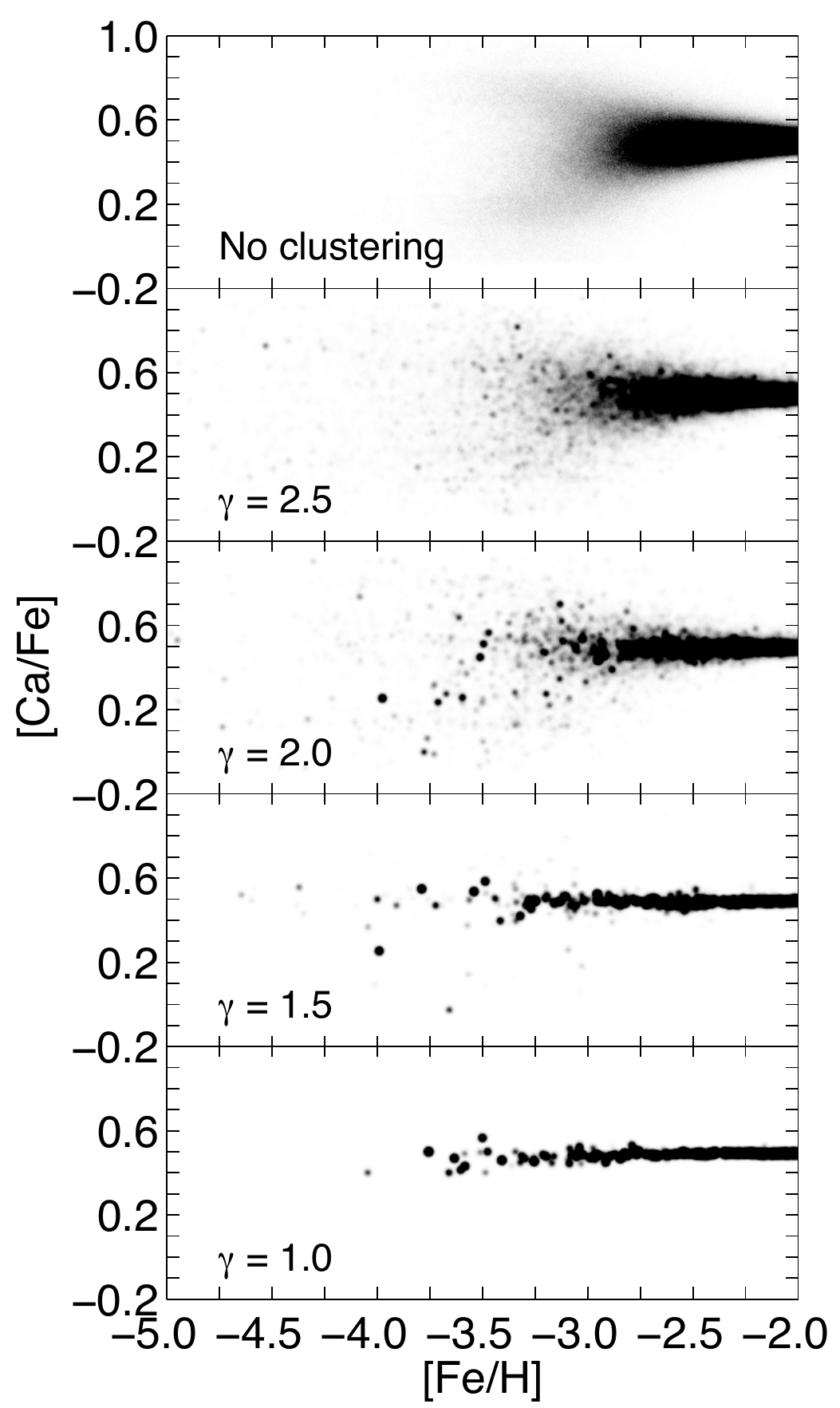}{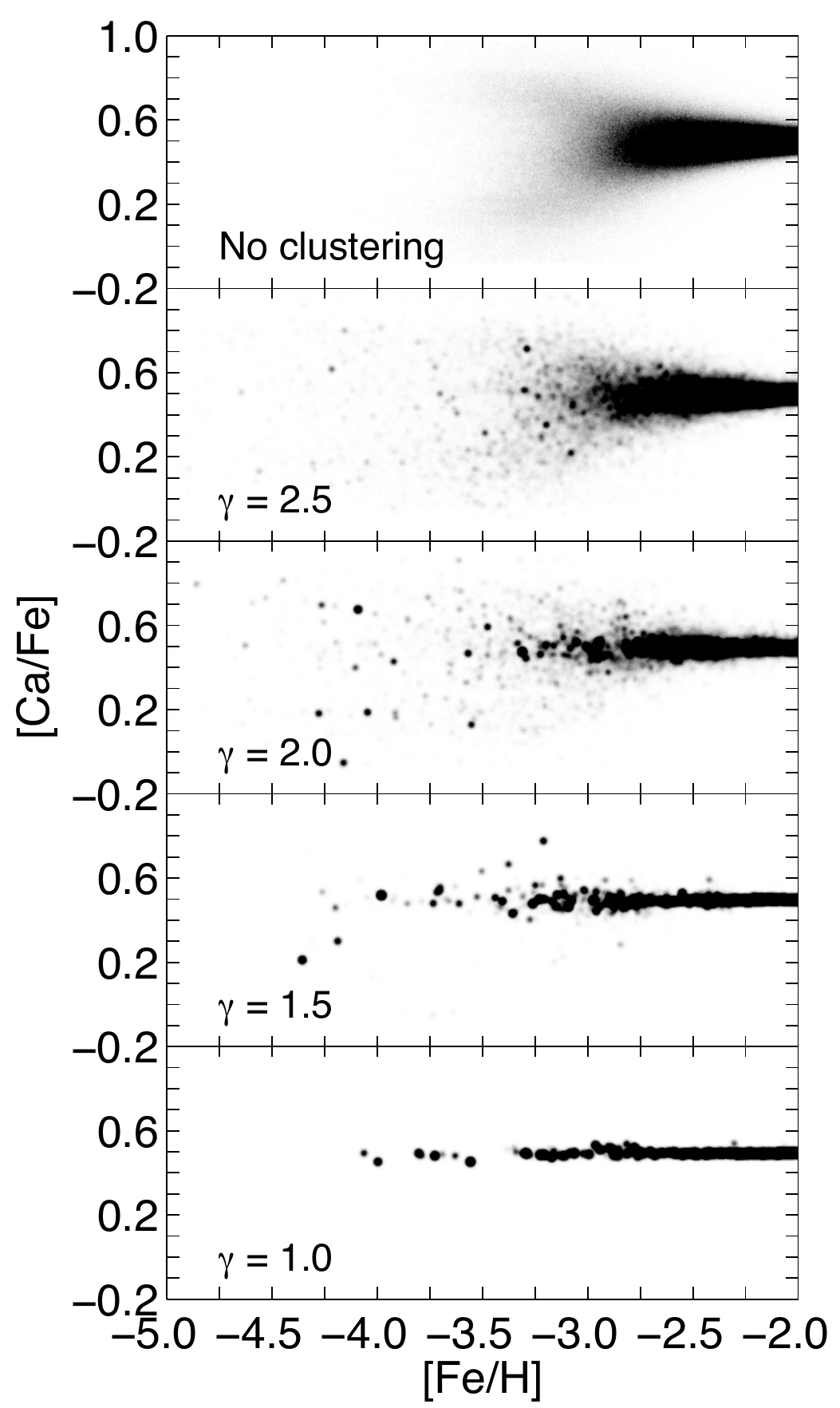}
\caption{ \label{CaFeH}
The results of the stochastic chemical evolution modelling in \S 5 for $\alpha$ element Ca compared to [Fe/H].
The five models from top to bottom are: (i) no clustering, (ii) $\gamma=2.5$, (iii) $\gamma=2.0$, (iv) $\gamma=1.5$,
(v) $\gamma=1.0$.  The left and right hand panels are two different realizations of the same 5 models. The 
clustering in abundance space becomes very apparent at low values of $\gamma$. High values of $\gamma$ are
barely distinguishable from the ``no clustering'' distribution in (i). Note also that the scattering
at a fixed value of [Fe/H] decreases dramatically with decreasing values of $\gamma$. The intrinsic dispersion
within individual clusters (0.01 dex) is much less than expected in real data; 
more realistic models with fewer data points and increased measurement error are presented in Fig.~\ref{noisy1}.
}
\end{figure}

The results of the clustered simulations for [Ca/Fe] and [r/Fe] with respect to [Fe/H]
are presented in Figs. \ref{CaFeH} and \ref{rFeH} respectively. For all models, the number of simulated
stars is $n_* \sim 10^7$ equivalent to a stellar mass of $3\times 10^6$ M$_\odot$ and a luminosity of
$10^6$ L$_\odot$ assuming a Salpeter IMF; the luminosity is two times higher
for a Kroupa IMF. To begin with, we do not consider the effect of distance on the observed
number counts; this is treated in \S 5.2. The modelled number of stars is much larger than we can expect to obtain
in nearby dwarfs, but the results provide sufficient resolution to understand the impact of
the ICMF parameters. The overall distribution of the stars with metallicity is shown in Fig.~\ref{MDF}.

There is a clear progression moving
from high $\gamma$ to low $\gamma$ in the occurrence of clustered abundance signatures.
The high $\gamma$ limit is indistinguishable from the beta (non-clustered) models. This is
not unexpected since, in the limit of $\gamma=2.5$, almost all stars are formed in small star clusters such that unique 
groupings in abundance space are poorly represented.

As $\gamma$ decreases, the clustering increases markedly but at the expense of the
dispersion in [X/Fe]. In the limit of $\gamma = 1$, the vertical spread has vanished over
all [Fe/H] with little or no abundance spread in [X/Fe]. While $\gamma = 1$ is smaller than what is 
observed, it is in fact a useful surrogate for demonstrating the impact of a higher maximum cloud mass
$M_{\rm max}$ for larger $\gamma$ values. For example, in Fig.~\ref{rFeHcut}, we show the
result of running $\gamma=1.5$ and $\gamma=2.0$ models with $M_{\rm max}\approx 10^6\;{\rm M}_\odot$. 
We see how the high mass cut-off at high $\gamma$ mimics the behaviour of
a lower value of $\gamma$: the vertical scatter is greatly reduced, and groupings
are seen to extend to lower [Fe/H].

In Fig.~\ref{IMF}, we show the cumulative fraction of stars and the integrated light as a function of 
stellar absolute magnitude for the model dwarf galaxy.
In Table 1, we give a rough breakdown of how many stars are expected for
the dwarf galaxy as a function of stellar apparent magnitude, distance and metallicity.
We consider V=16 and V=18 to be the bright and faint limit of an 8m class experiment; these
magnitude brackets become V=20 and V=22 on an ELT (Bland-Hawthorn et al 2010).

In the next section, we analyse these simulations using a group finder in order to provide an 
objective assessment of the amount of measurable clustering in the limit of high and low $n_*$
that we can expect. We apply this analysis for different [Fe/H] cut offs since, as we have seen,
the abundance plots become crowded as [Fe/H] increases.

\section{Clustering in ${\cal C}$-space: a statistical treatment}

\subsection{Group finding for large $n_*$}

In order to investigate the clustering in abundance
space (defined by [Fe/H] and [Ca/Fe] or [r/Fe] abundances), we use the density 
based hierarchical clustering algorithm EnLink \citep{sharma09}. 
The method is statistically objective and robust in its application; the method
is highly efficient and does not use pixellation or binning. 

EnLink is based on the fact that a system having more
than one group in a data set will have peaks and valleys in the density
distribution. A peak in a density distribution identifies a cluster and 
the set of points which can ``climb the peak," identified by following density 
gradients, are labeled as its members. The valleys in between the peaks 
represent intersections between the clusters. These are used to
define the boundaries of the clusters and also to form a parent-child 
relationship, thereby establishing a hierarchical organization of the 
groups. EnLink works by first calculating the densities of the data
points using a set of $q_{den}$ nearest neighbours and then organizes 
the points in groups by using nearest neighbour links. 

We present our results in the Appendix for both [Ca/Fe] and [r/Fe] for
different values of $\gamma$ over a wide dynamic range in the
observed number of data points.
We confirm that statistically significant groupings in abundance space
can be recovered, particularly at low values of $\gamma$. For this work,
we adopt a low cluster-mass limit of $M_{\rm min} = 5$ M$_\odot$ which 
is likely to be too conservative. For a fixed number of simulation particles,
decreasing $M_{\rm min}$ suppresses the number of detected groups.

\subsection{Group finding for small $n_*$}

Clustering should be present even in the limit of only a few data points.
To emphasize this fact, we have simulated the abundance measurements for
a dwarf galaxy at a distance of 30 kpc (see Figs.~\ref{noisy1} and \ref{noisy2})
as observed on 8m class and 30m class telescopes respectively.
We adopt a stellar mass of $3\times 10^5$ M$_\odot$ typical of
a faint dwarf galaxy. This object has about $10^6$ stars, a luminosity of $10^5$ 
L$_\odot$ and an absolute V mag of $M_V = -7.6$ assuming a Salpeter IMF.
The luminosity is a factor of two higher for a Kroupa IMF. The star counts are
consistent with the model values in Table 1 when scaled to the adopted lower mass.

For the 8m experiment (Fig.~\ref{noisy1}),
we assume measurement errors of 0.1 dex in both [r/Fe] and [Fe/H]. For 
$\gamma=1.5$ and $\gamma=2.0$, the effects of clustering are evident and
this holds true if we double the measurement error. The impact of this 
measurement uncertainty over the full simulation is shown in Fig.~\ref{rFeHsm}.
We conclude that, once more extensive studies are made of the nearest dwarf 
galaxies, the effects of clustering may become evident even before the advent
of ELTs. We envisage projects which focus on relatively few stars that appear
grouped in abundance space in low resolution spectroscopic data. Long
integrations and differential analysis will allow the null hypothesis to be
tested that the stars have identical abundances in all elements.

For the 30m experiment (Fig.~\ref{noisy2}),
we have assumed a general improvement in the atmospheric models and
the experimental errors, and therefore adopt errors of 0.05 dex. The effects of
clustering, which are easily seen, remain clearly visible even after a twofold
increase in the measurement errors in both axes. This simulation is a powerful
statement of the importance of multi-object echelles on ELTs.

For a fixed $q_{den}$ and data dimensionality,
the number of spurious groups due to Poisson noise increases linearly
with the total number of data points $n_*$. Therefore, EnLink can be used
in the limit of small $n_*$ with $q_{den}$ set to 3, i.e. a minimum of one
more than the number of data dimensions. But with so few data points,
the full power of EnLink is not being exploited such that more rudimentary
statistical techniques may be better. However, EnLink is particularly efficient
in treating data sets with more than two dimensions, even in the limit of 
small $n_*$.The significance of groups increases dramatically when we apply
EnLink on an abundance space with more dimensions. The normalizing
distribution in equation~\ref{eqn:beta} is easily extended to higher dimensions.
In Fig.~\ref{rcafe30}, we apply EnLink to the 3D ${\cal C}$-space 
([Fe/H], [Ca/Fe], [r/Fe]) for three different measurement errors (0.05, 0.1, 0.2 dex) 
typical of contemporary observations on 8m telescopes at V=18. Even in the 
presence of large errors, clustering signals are seen in all cases.

\section{Discussion}

This paper has explored the prospect of probing the mass scales of the first star clusters.
We stress that we have used a
lower cluster-mass limit of $M_{\rm min} = 5$ M$_\odot$ (typically 10 stars) which is very
conservative. A more reasonable value may be an order of magnitude higher (cf. 
Bland-Hawthorn et al 2010). For $\gamma=2$ and a fixed number of simulation particles,
the lower threshold has the effect of reducing cluster membership by a factor of ten, and
increasing the ``background'' by the same factor. This lowers the overall clustering 
signal by an order of magnitude. But we adopt the conservative lower mass limit in order
to account for a possible dwarf stellar population that are not born in clusters. We were unable
to find any observational constraints on the diffuse vs. clustered population in dwarf galaxies
(cf. Lada \& Lada 2003).

In our conservative analysis, we find that there is an intimate connection between properties of 
the ICMF and the amount of potentially detectable clustering in the abundance plane. 
A flat ICMF and/or a ICMF with a high mass cut-off produces strong clustering 
in the [Fe/H] vs. [X/Fe] abundance plane (${\cal C}$-space). While our models are 
inevitably oversimplified,
the phenomenon should be detectable on 8-10m telescopes (e.g. Figs.~\ref{noisy1}
and \ref{noisy2}). This gains support from existing observations of open clusters, globular
clusters and moving groups (Castro et al 1999; Shen et al 2005; Randich et al 2006; 
Sestito et al 2007; De Silva, Freeman \& Bland-Hawthorn 2009; Chou et al 2010; Bubar \& King
2010).  A clean ``clustering" signature in ${\cal C}$-space, particularly at low metallicity, 
is important for a 
number of reasons. First, it indicates the presence of massive star clusters in the early universe
and conceivably provides a constraint on the mass of the first systems.
Second, it provides a clean signal of the progenitor abundances in the cloud prior to cluster
formation. This abundance measurement is averaged over a substantial amount of gas and
is therefore not subject to mixing anomalies (Karlsson \& Gustafsson 2001) or mass transfer
in binaries (e.g. Suda et al 2004; Ryan et al 2005; Lucatello et al 2005).

Strong clustering would indicate a highly flattened ICMF, or a high mass cut-off. This could herald the 
onset of the formation of massive star clusters in dwarf galaxies (e.g. Bromm \& Clarke 2002). 
If the star formation efficiencies were 
low at that time, this may require supermassive gas clouds ($\gtrsim 10^7$ M$_\odot$) to have formed
even at the earliest times (Abel et al 2000), possibly consistent with the regular occurrence of massive star-forming
clumps at high redshift (Genzel et al 2006; Forster-Schreiber et al 2009; Elmegreen \& Elmegreen 2006).
Conversely, if such clustering was not observed, then we would infer that the slope of the
early ICMF is steep, or the maximum cluster size is relatively small compared to the
present day. But the observed scatter would need to be consistent with the non-detection
of clustering.

In future investigations, from the perspective of chemical signatures, 
we will look at the degeneracy between steep ICMFs with a high mass
cut-off and flatter ICMFs with a lower mass cut-off (see \S 4.2). We will also look at
a wider class of chemical elements that are considered in numerical simulations of the first stars.
We will look at the improvement in cluster identification with more chemical 
elements, particularly in the limit of small $n_*$. The clustered abundance signatures will provide unique 
insight into the most ancient star clusters. These signatures are 
signposts of the chemistry immediately before the onset of star formation, untainted by mass 
transfer in close binary pairs or incomplete mixing anomalies.

Finally, it is an extraordinary fact that we can probe back to the first billion years
from observations of the {\it local} universe. We can say with absolute certainty that stars existed 
at this time. These were responsible for the first chemical elements (Ryan-Weber, Pettini \& Madau 2006)
and for reionizing the fog of hydrogen that permeated the early Universe (Fan et al 2002). 
Precisely when the first star clusters formed is unknown. It seems likely, however, that
gas was able to fragment at very high density even at primordial abundance levels (Clark et al 2008).
It may be possible to directly probe these environments in an era of the Atacama Large Millimetre Array and the 
James Webb Space Telescope. But we believe that some of the most important insights, particularly with
regard to progenitor yields, will undoubtedly come from 
near-field cosmology. To this end, it will be necessary to equip the next generation of ELTs with wide-field
multi-object spectrographs that operate at high spectroscopic resolution (R$\gtrsim$20,000).

\acknowledgments  JBH is supported by an Federation Fellowship from the Australian Research Council (ARC),
which also funds TK's research position. SS is funded under the ARC DP grant 0988751 which partially supports the
HERMES project.
MRK is supported by the National Science Foundation through grant AST-0807739; by NASA through the Spitzer 
Space Telescope Theoretical Research Program, provided by a contract issued by the Jet Propulsion Laboratory; 
and by the Alfred P. Sloan Foundation through a Sloan Research Fellowship. JBH is indebted to Merton College,
Oxford for a Visiting Research Fellowship, and to the Leverhulme Trust for a Visiting Professorship to Oxford. 
JBH and TK are grateful to the BIPAC Institute, Oxford for their hospitality.

\newpage
\begin{table}[htdp]
\caption{ \label{tab1}
Log of the expected numbers of stars as a function of V mag for an old,
metal poor dwarf spheroidal (see Fig.~\ref{MDF})
with 10$^7$ stars (stellar mass $\approx$ $3\times 10^6$ M$_\odot$) 
at distances of 10, 30 and 100 kpc (col. 1). Columns 2-5 are the star
counts for a Salpeter IMF in 4 metallicity bins; columns 6-9 are the counts
for a Kroupa IMF in the same bins determined from the MDF 
in Fig. 2. The 4 metallicity bins are [Fe/H] = (-5:-4), (-4:-3), (-3:-2), (-2:-1);
dashes indicate that no stars are expected.}
\begin{center}
\begin{tabular}{rccccccccccc}
 & & \multicolumn{4}{c}{Salpeter} & & \multicolumn{4}{c}{Kroupa} \\ \cline{3-6} \cline{8-11}
 & & & & & & & & & & \\
 $[$Fe/H$]$ & & -5:-4 & -4:-3 & -3:-2 & -2:-1 & & -5:-4 & -4:-3 & -3:-2 & -2:-1 \\
 & & & & & & & & & & \\
 V = 16 & & & & & & & & & & \\ \cline{1-1}
 & & & & & & & & & & \\
  10 & &  0.7 &  1.9 &  3.0 &  3.5 &  &  1.0 &  2.2 &  3.3 &  3.8   \\
  30 & & - &  0.8 &  1.9 &  2.4 &  & - &  1.1 &  2.2 &  2.7   \\
 100 & & - & - &  0.4 &  1.0 &  & - & - &  0.7 &  1.3  \\
 & & & & & & & & & & \\

V = 18 & & & & & & & & & & \\ \cline{1-1}
   & & & & & & & & & & \\
 10 & &  1.1 &  2.3 &  3.4 &  3.9 &  &  1.4 &  2.6 &  3.7 &  4.2  \\
  30 & &  0.7 &  1.9 &  3.0 &  3.5 &  &  0.9 &  2.2 &  3.2 &  3.8  \\
 100 & & - &  0.6 &  1.7 &  2.2 &  & - &  0.9 &  2.0 &  2.5  \\
 & & & & & & & & & & \\

V = 20 & & & & & & & & & & \\ \cline{1-1}
  & & & & & & & & & & \\
 10 & &  2.2 &  3.4 &  4.5 &  5.0 &  &  2.5 &  3.7 &  4.8 &  5.3   \\
  30 & &  1.0 &  2.2 &  3.3 &  3.8 &  &  1.3 &  2.5 &  3.6 &  4.1   \\
 100 & &  0.1 &  1.3 &  2.4 &  2.9 &  &  0.4 &  1.6 &  2.7 &  3.2  \\
 & & & & & & & & & & \\

V = 22 & & & & & & & & & & \\ \cline{1-1}
   & & & & & & & & & & \\
 10 & &  2.8 &  4.0 &  5.1 &  5.6 &  &  3.0 &  4.3 &  5.3 &  5.8   \\
  30 & &  2.0 &  3.3 &  4.3 &  4.8 &  &  2.3 &  3.5 &  4.6 &  5.1   \\
 100 & &  0.9 &  2.1 &  3.2 &  3.7 &  &  1.2 &  2.4 &  3.4 &  4.0  \\
& & & & & & & & & & \\  \hline \hline

\end{tabular}
\end{center} 
\label{default}
\end{table}

\newpage

\bibliographystyle{apj}

\begin{thebibliography}{}
\bibitem[Abel et al.(2000)]{abel00} Abel, T., Bryan, G.~L., \& Norman, M.~L.\ 2000, \apj, 540, 39 
\bibitem[Abel et al.(2002)]{abel02} Abel, T., Bryan, G.~L., \& Norman, M.~L.\ 2002, Science, 295, 93 
\bibitem[Andr\'e et al (2007)]{andre07} Andr\'e, P., Belloche, A., Motte, F., \& Peretto, N. 2007, A\&A, 472, 519
\bibitem[Argast et al.(2000)]{argast00} Argast, D., Samland, M., Gerhard, O.~E., \& Thielemann, F.-K.\ 2000, \aap, 356, 873 
\bibitem[Argast et al.(2004)]{argast04} Argast, D., Samland, M., Thielemann, F.-K., \& Qian, Y.-Z.\ 2004, \aap, 416, 997 
\bibitem[Ascasibar \& Binney(2005)]{ascasibar05} Ascasibar, Y., \& Binney, J.\ 2005, \mnras, 356, 872 
\bibitem[Audouze \& Silk(1995)]{audouze95} Audouze, J., \& Silk, J.\ 1995, \apjl, 451, L49 
\bibitem[Ballero et al.(2006)]{ballero06} Ballero, S.~K., Matteucci, F., \& Chiappini, C.\ 2006, New Astronomy, 11, 306 
\bibitem[Beers \& Christlieb(2005)]{beers05} Beers, T.~C., \& Christlieb, N.\ 2005, \araa, 43, 531 
\bibitem[Belokurov et al.(2009)]{belokurov09} Belokurov, V., et al.\ 2009, \mnras, 397, 1748 
\bibitem[Bland-Hawthorn \& Freeman(2004)]{bland04} Bland-Hawthorn, J., \& Freeman, K.~C.\ 2004, Publications of the Astronomical Society of Australia, 21, 110 
\bibitem[Bland-Hawthorn \& Peebles(2006)]{bland06} Bland-Hawthorn, J., \& Peebles, P.~J.~E.\ 2006, Science, 313, 311 
\bibitem[Bland-Hawthorn et al.(2010)]{bland10} Bland-Hawthorn, J., Krumholz, M., \& Freeman, K.\ 2010, \apj, 713, 166
\bibitem[Bromm \& Clarke(2002)]{bromm02a} Bromm, V., \& Clarke, C.~J.\ 2002, \apjl, 566, L1 
\bibitem[Bromm et al.(2002)]{bromm02b} Bromm, V., Coppi, P.~S., \& Larson, R.~B.\ 2002, \apj, 564, 23 
\bibitem[Brook et al.(2007)]{brook07} Brook, C.~B., Kawata, D., Scannapieco, E., Martel, H., \& Gibson, B.~K.\ 2007, \apj, 661, 10 
\bibitem[Bubar \& King(2010)]{bubar10} Bubar, E.~J., \& King, J.~R.\ 2010, \aj, 140, 293 
\bibitem[Castro et al.(1999)]{castro99} Castro, S., Porto de Mello, G.~F., \& da Silva, L.\ 1999, \mnras, 305, 693 
\bibitem[Cayrel et al.(2004)]{cayrel04} Cayrel, R., et al.\ 2004, \aap, 416, 1117 
\bibitem[Chou et al.(2010)]{chou10} Chou, M.-Y., et al. 2010, \apj, 708, 1290
\bibitem[Christlieb et al.(2002)]{christlieb02} Christlieb, N., et al.\ 2002, \nat, 419, 904 
\bibitem[Cioffi et al.(1988)]{cioffi88} Cioffi, D.~F., McKee, C.~F., \& Bertschinger, E.\ 1988, \apj, 334, 252
\bibitem[Clark et al (2008)]{clark08} Clark, P.~C., Glover, S.~C.~O., \& Klessen, R.~S. 2008, ApJ, 672, 757
\bibitem[Clark et al (2009)]{clark09} Clark, P.~C., Glover, S.~C.~O., Bonnell, I.~A., \& Klessen, R.~S. 2009, ApJ, submitted, arXiv:0904.3302
\bibitem[Cohen et al.(2007)]{cohen07} Cohen, J.~G., McWilliam, A., Christlieb, N., Shectman, S., Thompson, I., Melendez, J., Wisotzki, L., \& Reimers, D.\ 2007, \apjl, 659, L161 
\bibitem[Coleman et al.(2004)]{coleman04} Coleman, M., Da Costa, G.~S., Bland-Hawthorn, J., Mart{\'{\i}}nez-Delgado, D., Freeman, K.~C., \& Malin, D.\ 2004, \aj, 127, 832 
\bibitem[De Silva et al. (2006)]{desilva06} De Silva, G.~M., Sneden, C., Paulson, D.~B., Asplund, M., Bland-Hawthorn, J., Bessell, M.~S., \& Freeman, K.~C.\ 2006, \aj, 131, 455 
\bibitem[De Silva et al. (2007a)]{desilva07a} De Silva, G.~M., Freeman, K.~C., Bland-Hawthorn, J., Asplund, M., \& Bessell, M.~S.\ 2007, \aj, 133, 694
\bibitem[De Silva et al. (2007b)]{desilva07b} De Silva, G.~M., Freeman, K.~C., Asplund, M., Bland-Hawthorn, J., Bessell, M.~S., \& Collet, R.\ 2007, \aj, 133, 1161 
\bibitem[De Silva et al.(2009)]{desilva2009} De Silva, G.~M., Freeman, K.~C., \& Bland-Hawthorn, J. 2009, PASA, 26, 11
\bibitem[Elmegreen \& Elmegreen(2006)]{elmegreen06} Elmegreen, B.~G., \& Elmegreen, D.~M.\ 2006, \apj, 650, 644 
\bibitem[Elmegreen(2010)]{elmegreen10} Elmegreen, B.~G.\ 2010, arXiv:1003.0798 
\bibitem[Escala \& Larson(2008)]{escala08} Escala, A., \& Larson, R.~B.\ 2008, \apjl, 685, L31 
\bibitem[Fall et al.(2005)]{fall05a} Fall, S.~M., Chandar, R., \& Whitmore, B.~C.\ 2005, \apjl, 631, L133 
\bibitem[Fall et al.(2009)]{fall09a} Fall, S.~M., Chandar, R., \& Whitmore, B.~C.\ 2009, \apj, 704, 453 
\bibitem[Fall et al.(2010)]{fall10a} Fall, S.~M., Krumholz, M.~R., \& Matzner, C.~D. 2010, \apjl, 710, L142
\bibitem[Fan et al.(2002)]{fan02} Fan, X., Narayanan, V.~K., Strauss, M.~A., White, R.~L., Becker, R.~H., Pentericci, L., \& Rix, H.-W.\ 2002, \aj, 123, 1247 
\bibitem[Feltzing et al.(2009)]{feltzing09} Feltzing, S., Eriksson, K., Kleyna, J., \& Wilkinson, M.~I.\ 2009, \aap, 508, L1 
\bibitem[Fields et al.(2002)]{fields02} Fields, B.~D., Truran, J.~W., \& Cowan, J.~J.\ 2002, \apj, 575, 845 
\bibitem[F{\"o}rster Schreiber et al.(2009)]{forster09} F{\"o}rster Schreiber, N.~M., et al.\ 2009, \apj, 706, 1364 
\bibitem[Frebel et al.(2005)]{frebel05} Frebel, A., et al.\ 2005, \nat, 434, 871 
\bibitem[Frebel et al.(2008)]{frebel08} Frebel, A., Collet, R., Eriksson, K., Christlieb, N., \& Aoki, W.\ 2008, \apj, 684, 588 
\bibitem[Frebel et al.(2010)]{frebel10a} Frebel, A., Simon, J.~D., Geha, M., \& Willman, B.\ 2010, \apj, 708, 560 
\bibitem[Frebel et al.(2010)]{frebel10b} Frebel, A., Kirby, E.~N., \& Simon, J.~D.\ 2010, \nat, 464, 72 
\bibitem[Freeman \& Bland-Hawthorn(2002)]{fbh02} Freeman, K. \& Bland-Hawthorn, J. 2002, ARA\&A, 40, 487
\bibitem[Frinchaboy et al.(2008)]{frinchaboy08} Frinchaboy, P.~M., Marino, A.~F., Villanova, S., Carraro, G., Majewski, S.~R., \& Geisler, D.\ 2008, \mnras, 391, 39 
\bibitem[Fulbright et al.(2004)]{fulbright04} Fulbright, J.~P., Rich, R.~M., \& Castro, S.\ 2004, \apj, 612, 447 
\bibitem[Genzel et al (2006)]{genzel06} Genzel, R., et al. 2006, Nature, 442, 786
\bibitem[Gratton et al.(2004)]{gratton04a} Gratton, R., Sneden, C. \& Carretta, E. 2004, \araa, 42, 385
\bibitem[Helmi(2008)]{helmi08} Helmi, A.\ 2008, \aapr, 15, 145 
\bibitem[Hennebelle \& Chabrier (2008)]{hennebelle08} Hennebelle, P., \& Chabrier, G. 2008, ApJ, 684, 395
\bibitem[Heyer \& Brunt (2004)]{heyer04} Heyer, M.~H., \& Brunt, C.~M.\ 2004, \apjl, 615, L45 
\bibitem[Ibata et al.(1995)]{ibata95} Ibata, R.~A., Gilmore, G., \& Irwin, M.~J.\ 1995, \mnras, 277, 781 
\bibitem[Ishimaru \& Wanajo(1999)]{ishimura99} Ishimaru, Y., \& Wanajo, S.\ 1999, \apjl, 511, L33 
\bibitem[Izotov \& Thuan(2004)]{izotov04} Izotov, Y.~I., \& Thuan, T.~X.\ 2004, \apj, 616, 768 
\bibitem[Joggerst et al.(2010)]{joggerst10} Joggerst, C.~C., Almgren, A., Bell, J., Heger, A., Whalen, D., \& Woosley, S.~E.\ 2010, \apj, 709, 11 
\bibitem[Karlsson(2005)]{karlsson05a} Karlsson, T.\ 2005, \aap, 439, 93 
\bibitem[Karlsson(2006)]{karlsson06} Karlsson, T.\ 2006, \apjl, 641, L41 \bibitem[Karlsson \& Gustafsson(2001)]{2001A&A...379..461K} Karlsson, T., \& Gustafsson, B.\ 2001, \aap, 379, 461 
\bibitem[Karlsson \& Gustafsson(2005)]{karlsson05b} Karlsson, T., \& Gustafsson, B.\ 2005, \aap, 436, 879 
\bibitem[Karlsson et al.(2008)]{karlsson08} Karlsson, T., Johnson, J.~L., \& Bromm, V.\ 2008, \apj, 679, 6 
\bibitem[Kirby et al.(2008)]{kirby08} Kirby, E.~N., Simon, J.~D., Geha, M., Guhathakurta, P., \& Frebel, A.\ 2008, \apjl, 685, L43 
\bibitem[Koch et al.(2008)]{koch08} Koch, A., McWilliam, A., Grebel, E.~K., Zucker, D.~B., \& Belokurov, V.\ 2008, \apjl, 688, L13 
\bibitem[Kroupa \& Boily(2002)]{kroupa02} Kroupa, P., \& Boily, C.~M.\ 2002, \mnras, 336, 1188 
\bibitem[Krumholz et al. (2005)]{krumholz05} Krumholz, M.~R., McKee, C.~F., \& Klein, R.~I. 2005, Nature, 438, 332
\bibitem[Lada \& Lada(2003)]{lada03} Lada, C.~J., \& Lada, E.~A.\ 2003, \araa, 41, 57
\bibitem[Larsen(2009)]{larsen09} Larsen, S.~S.\ 2009, \aap, 503, 467 
\bibitem[Lucatello et al.(2005)]{lucatello05} Lucatello, S., Tsangarides, S., Beers, T.~C., Carretta, E., Gratton, R.~G., \& Ryan, S.~G.\ 2005, \apj, 625, 825 
\bibitem[Maraston et al.(2004)]{maraston04} Maraston, C., Bastian, N., Saglia, R.~P., Kissler-Patig, M., Schweizer, F., \& Goudfrooij, P.\ 2004, \aap, 416, 467 
\bibitem[Mateo(1998)]{mateo98} Mateo, M.~L.\ 1998, \araa, 36, 435 
\bibitem[Maund et al.(2009)]{maund09} Maund, J.~R., Wheeler, J.~C., Baade, D., Patat, F., H{\"o}flich, P., Wang, L., \& Clocchiatti, A.\ 2009, \apj, 705, 1139 
\bibitem[McKee \& Ostriker(2007)]{mckee07b} McKee, C.~F., \& Ostriker, E.~C.\ 2007, \araa, 45, 565 
\bibitem[McWilliam et al.(2009)]{mcwilliam09} McWilliam, A., Simon, J.~D., \& Frebel, A.\ 2009, astro2010: The Astronomy and Astrophysics Decadal Survey, 2010, 200 
\bibitem[McWilliam(1997)]{mcwilliam97} McWilliam, A.\ 1997, \araa, 35, 503 
\bibitem[Murray \& Lin(1990)]{murray90} Murray, S.~D., \& Lin, D.~N.~C.\ 1990, \apj, 363, 50 
\bibitem[Nomoto et al.(2005)]{nomoto05} Nomoto, K., Tominaga, N., Umeda, H., Maeda, K., Ohkubo, T., Deng, J., \& Mazzali, P.~A.\ 2005, The Fate of the Most Massive Stars, 332, 374 
\bibitem[Nomoto et al.(2006)]{nomoto06} Nomoto, K., Tominaga, N., Umeda, H., Kobayashi, C., \& Maeda, K.\ 2006, Nuclear Physics A, 777, 424 
\bibitem[Okrochkov \& Tumlinson(2010)]{okrochkov10} Okrochkov, M., \& Tumlinson, J.\ 2010, \apjl, 716, L41
\bibitem[Padoan \& Nordlund (2002)]{padoan02} Padoan, P., \& Nordlund, \AA. 2002, ApJ, 576, 870
\bibitem[Portegies Zwart et al.(2010)]{portegies10} Portegies Zwart, S., McMillan, S., \& Gieles, M.\ 2010, arXiv:1002.1961 
\bibitem[Qian(2000)]{qian00} Qian, Y.-Z.\ 2000, \apjl, 534, L67 
\bibitem[Qian(2001)]{qian01} Qian, Y.-Z.\ 2001, \apjl, 552, L117 
\bibitem[Raiteri et al.(1999)]{raiteri99} Raiteri, C.~M., Villata, M., Gallino, R., Busso, M., \& Cravanzola, A.\ 1999, \apjl, 518, L91 
\bibitem[Randich et al.(2006)]{randich06} Randich, S., Sestito, P., Primas, F., Pallavicini, R., \& Pasquini, L.\ 2006, \aap, 450, 557 
\bibitem[Roederer et al.(2009)]{roederer09} Roederer, I.~U., Kratz, K.-L., Frebel, A., Christlieb, N., Pfeiffer, B., Cowan, J.~J., \& Sneden, C.\ 2009, \apj, 698, 1963 
\bibitem[Ryan et al.(1996)]{ryan96} Ryan, S.~G., Norris, J.~E., \& Beers, T.~C.\ 1996, \apj, 471, 254 
\bibitem[Ryan et al.(2005)]{ryan05} Ryan, S.~G., Aoki, W., Norris, J.~E., \& Beers, T.~C.\ 2005, \apj, 635, 349 
\bibitem[Ryan-Weber et al.(2006)]{ryanweber06} Ryan-Weber, E.~V., Pettini, M., \& Madau, P.\ 2006, \mnras, 371, L78 
\bibitem[Scannapieco et al.(2006)]{scannapieco06} Scannapieco, E., Kawata, D., Brook, C.~B., Schneider, R., Ferrara, A., 
\& Gibson, B.~K.\ 2006, \apj, 653, 285
\bibitem[Schneider \& Omukai(2010)]{schneider10} Schneider, R., \& Omukai, K.\ 2010, \mnras, 402, 429 
\bibitem[Sestito et al.(2007)]{sestito07} Sestito, P., Randich, S., \& Bragaglia, A.\ 2007, \aap, 465, 185 
\bibitem[Sharma \& Johnston(2009)]{sharma09}{Sharma}, S. \& {Johnston}, K.~V. 2009, \apj, 703, 1061
\bibitem[Shen et al.(2005)]{shen05} Shen, Z.-X., Jones, B., Lin, D.~N.~C., Liu, X.-W., \& Li, S.-L.\ 2005, \apj, 635, 608 
\bibitem[Shigeyama \& Tsujimoto(1998)]{shigeyama98} Shigeyama, T., \& Tsujimoto, T.\ 1998, \apjl, 507, L135 
\bibitem[Simon \& Geha(2007)]{simon07} Simon, J.~D., \& Geha, M.\ 2007, \apj, 670, 313 
\bibitem[Suda et al.(2004)]{suda04} Suda, T., Aikawa, M., Machida, M.~N., Fujimoto, M.~Y., \& Iben, I., Jr.\ 2004, \apj, 611, 476 
\bibitem[Tan et al. (2006)]{tan06a} Tan, J.~C., Krumholz, M.~R., \& McKee, C.~F.\ 2006, \apjl, 641, L121 
\bibitem[Tolstoy et al.(2009)]{tolstoy09} Tolstoy, E., Hill, V., \& Tosi, M.\ 2009, \araa, 47, 371 
\bibitem[Toomre (1964)]{toomre64} Toomre, A. 1964, \apj, 139, 1217
\bibitem[Travaglio et al.(2001)]{travaglio01} Travaglio, C., Galli, D., \& Burkert, A.\ 2001, \apj, 547, 217 
\bibitem[Tsuribe \& Omukai(2006)]{tsuribe06} Tsuribe, T., \& Omukai, K.\ 2006, \apjl, 642, L61 
\bibitem[Tsuribe \& Omukai(2008)]{tsuribe08} Tsuribe, T., \& Omukai, K.\ 2008, \apjl, 676, L45 
\bibitem[Tumlinson(2010)]{tumlinson10a} Tumlinson, J, 2010, \apj, 708, 1398
\bibitem[Venn et al.(2004)]{venn04} Venn, K.~A., Irwin, M., Shetrone, M.~D., Tout, C.~A., Hill, V., \& Tolstoy, E.\ 2004, \aj, 128, 1177
\bibitem[Villanova et al.(2010)]{villanova10} Villanova, S., Randich, S., Geisler, D., Carraro, G., \& Costa, E.\ 2010, \aap, 509, A102 
\bibitem[Walsh et al.(2007)]{walsh07} Walsh, S.~M., Jerjen, H., \& Willman, B.\ 2007, \apjl, 662, L83 
\bibitem[Wang \& Wheeler(2008)]{wang08} Wang, L., \& Wheeler, J.~C.\ 2008, \araa, 46, 433 
\bibitem[Wasserburg \& Qian(2000)]{wasserburg00} Wasserburg, G.~J., \& Qian, Y.-Z.\ 2000, \apjl, 538, L99 
\bibitem[White \& Springel(2000)]{white00} White, S.~D.~M., \& Springel, V.\ 2000, The First Stars, 327
\bibitem[Yamaguchi et al.(1999)]{yamaguchi99} Yamaguchi, R., Saito, H., Mizuno, N., Mine, Y., Mizuno, A., Ogawa, H., \& Fukui, Y.\ 1999, \pasj, 51, 791 
\bibitem[Yamaguchi et al.(2001)]{yamaguchi01} Yamaguchi, R., Mizuno, N., Onishi, T., Mizuno, A., \& Fukui, Y.\ 2001, \apjl, 553, L185 
\end{thebibliography}

\begin{figure}
\plottwo{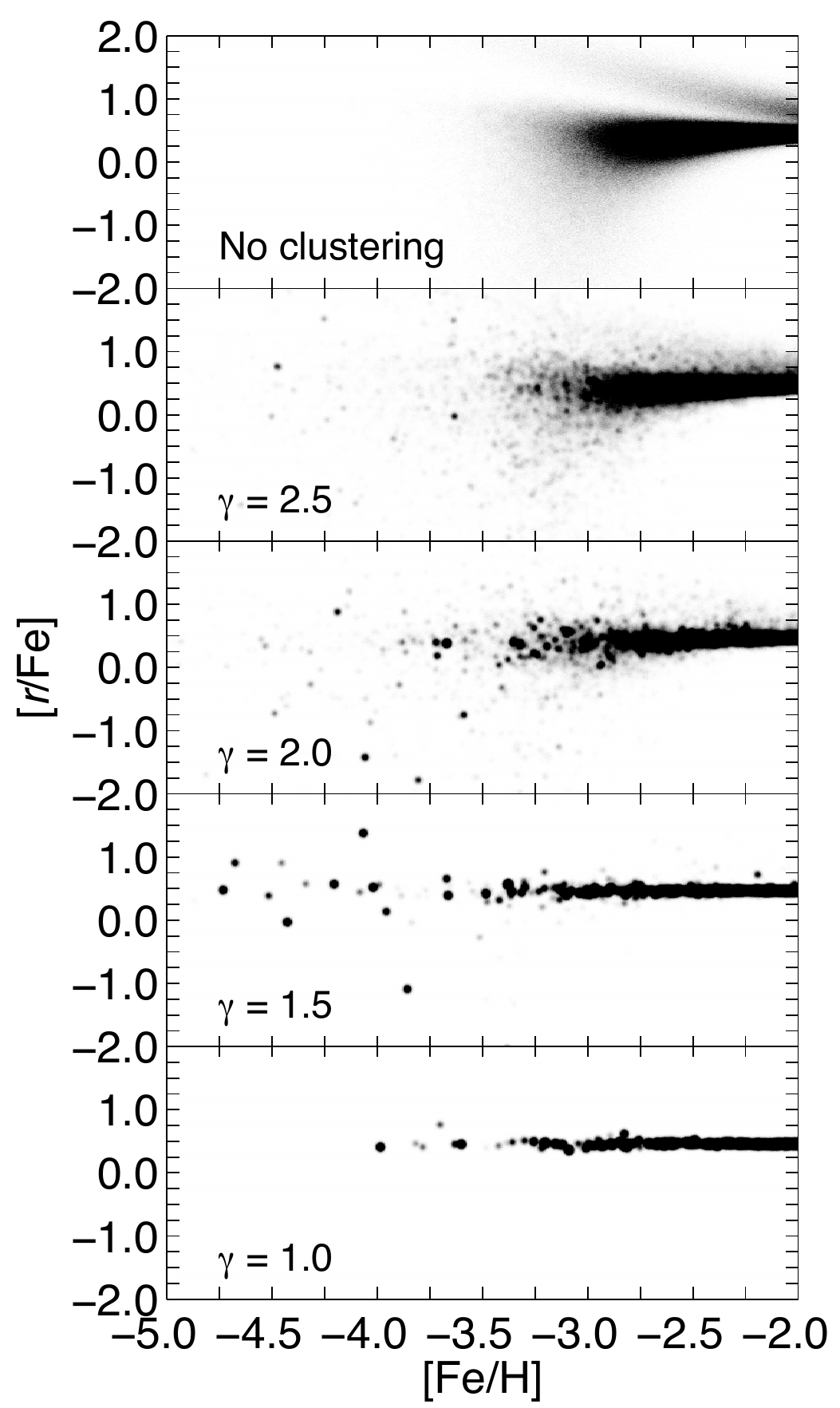}{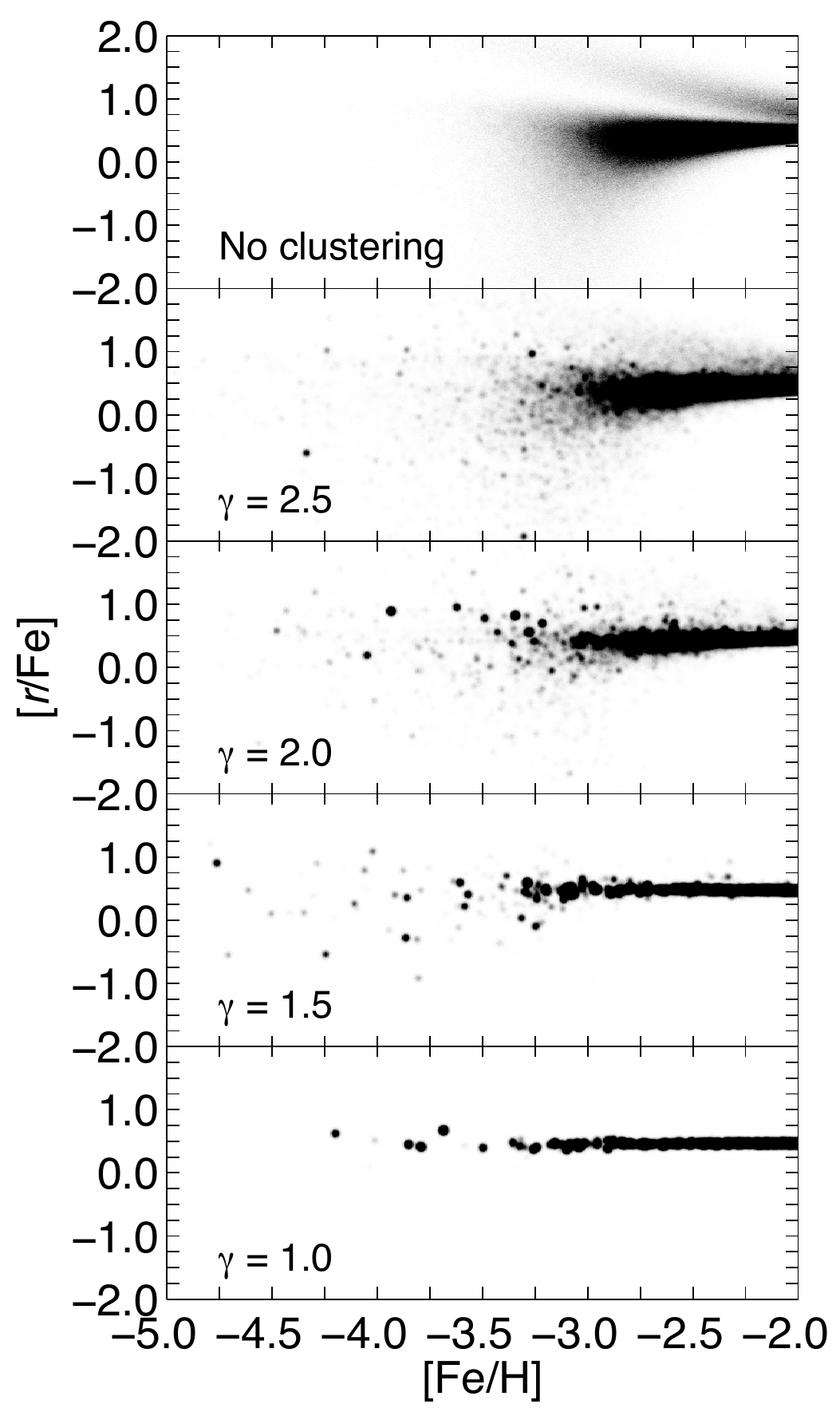}
\caption{ \label{rFeH}
The results of the stochastic chemical evolution modelling in \S 5 for r-process element (specifically Eu) compared to [Fe/H].
The vertical extent is 4 dex in [r/Fe], i.e. a fourfold increase over [Ca/Fe] in Fig.~\ref{CaFeH}. (The vertical and
horizontal axes are presented with the correct aspect ratio in Figs.~\ref{rFeHcut} and \ref{rFeHsm}.)
The five models from top to bottom are: (i) no clustering, (ii) $\gamma=2.5$, (iii) $\gamma=2.0$, (iv) $\gamma=1.5$,
(v) $\gamma=1.0$.  The left and right hand panels are two different realizations of the same 5 models. The 
clustering in abundance space becomes very apparent at low values of $\gamma$. High values of $\gamma$ are
barely distinguishable from the ``no clustering'' distribution in (i). Note also that the scattering
at a fixed value of [Fe/H] decreases dramatically with decreasing values of $\gamma$. In order to make clustered
points more circular, the intrinsic dispersion 
within individual clusters is $\Delta$[Fe/H]=0.01 dex and $\Delta$[Eu/Fe]=0.035 dex. }
\end{figure}

\begin{figure}
\centering \includegraphics[width=1.0\textwidth]{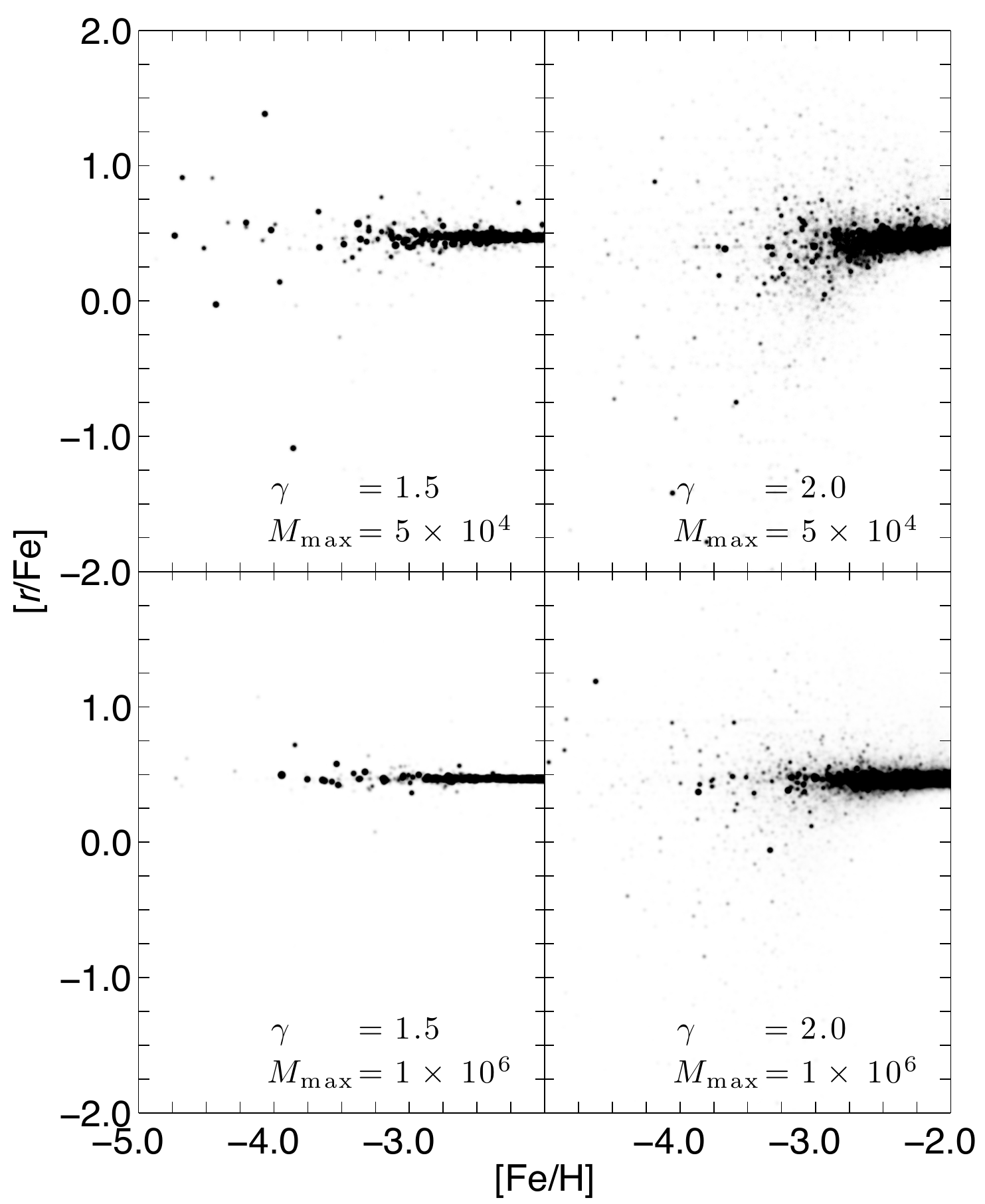}
\caption{ \label{rFeHcut}
The results of the stochastic chemical evolution modelling in \S 5 for the 2D space ${\cal C}$([Fe/H], [r/Fe]).
The top two panels are repeated from Fig.~\ref{rFeH} for which $\gamma=1.5$ (left) and $\gamma=2.0$ (right),
both with a high mass cut-off $M_{\rm max} = 5.0 \times 10^4$ M$_\odot$. The bottom two panels use $\gamma=1.5$
(left) and $\gamma=2.0$ (right) but with a high mass cut-off $M_{\rm max} = 1.0 \times 10^6$ M$_\odot$, twenty times higher 
than was used for the upper panels. Note how the high mass cut-off at high $\gamma$ mimics the behaviour of
a lower value of $\gamma$: the distribution is everywhere flattened in the vertical direction, and groupings
are seen to extend to lower [Fe/H]. This behaviour is particularly apparent in the left-hand figures.
}
\end{figure}

\begin{figure}
\centering \includegraphics[width=1.0\textwidth]{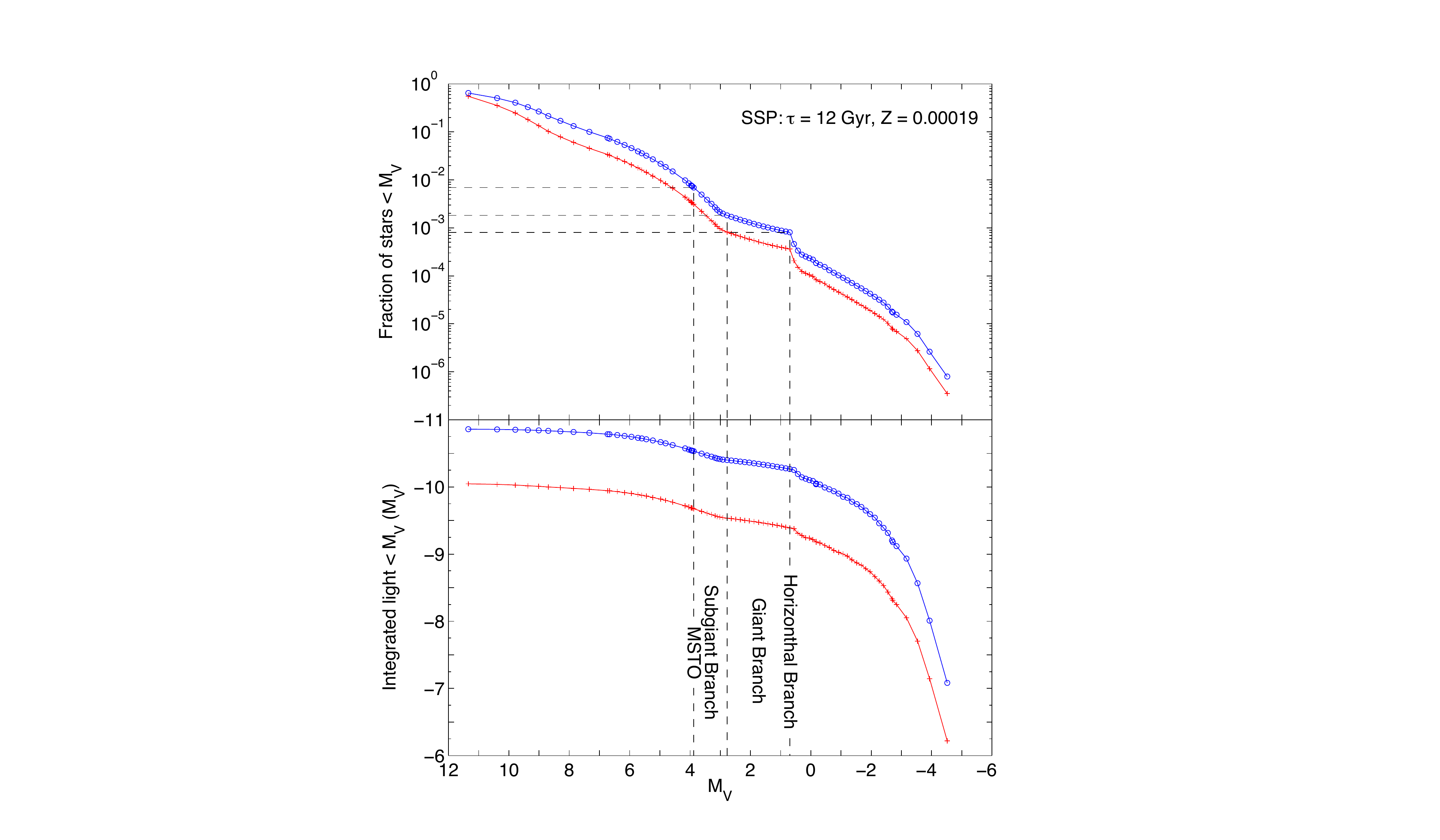}
\caption{ \label{IMF}
(Top) Cumulative fraction of stars brighter than a given stellar absolute magnitude $M_{\rm V}$ for a Salpeter
(lower) and Kroupa (upper) IMF. The high and low stellar mass cut-offs for both IMFs are 0.1 M$_\odot$ and 100 
M$_\odot$. The stellar population has a mass of $3\times 10^6$ M$_\odot$ and we use the 
Padova isochrones (see text) for a 12 Gyr old single burst, metal poor population. (Bottom)
Integrated light from this same population brighter than a given stellar absolute magnitude $M_{\rm V}$.
}
\end{figure}

\begin{figure}
\plotone{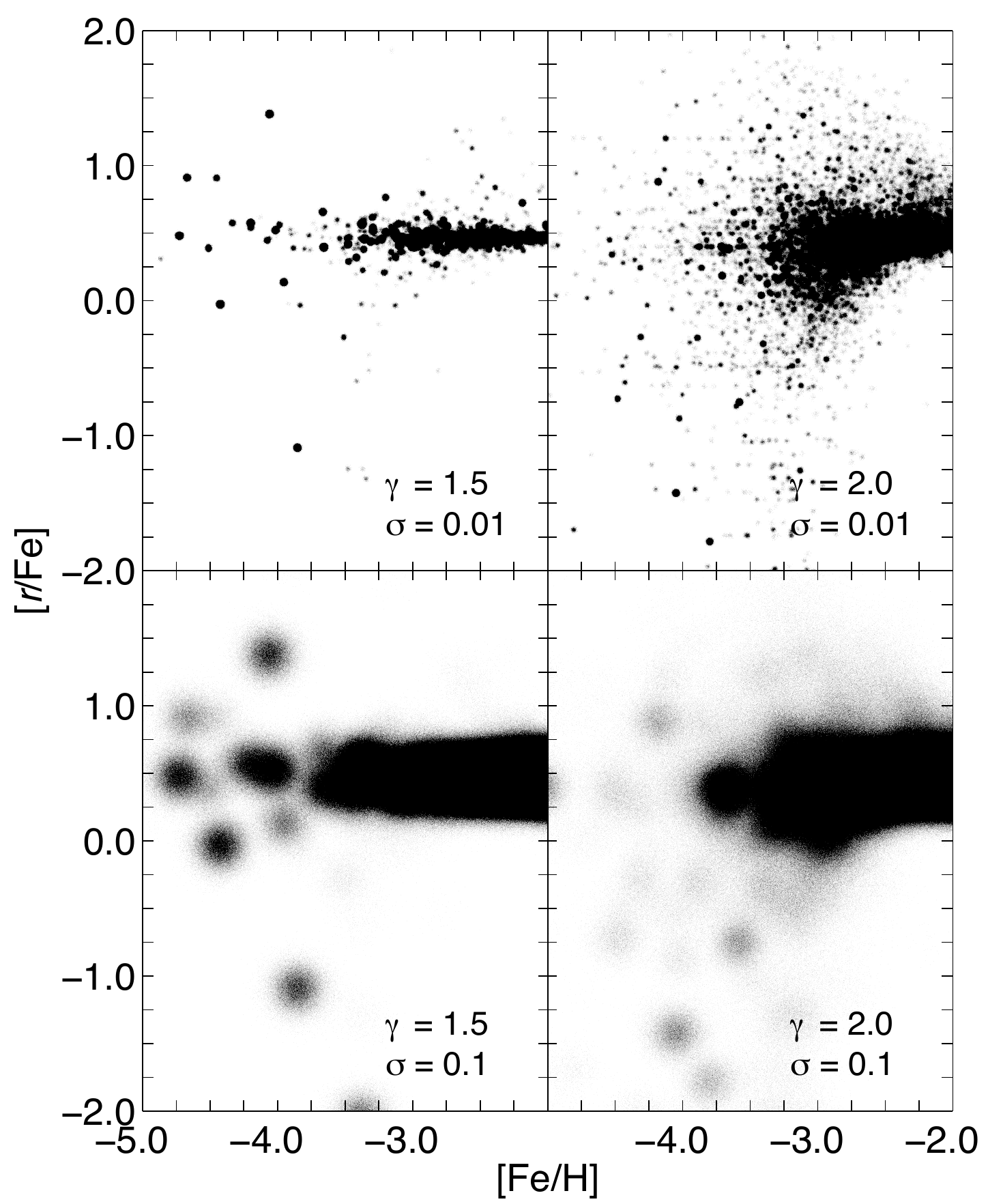}
\caption{ \label{rFeHsm}
The same models presented in Fig.~\ref{rFeH} but without the fourfold compression in
the [r/Fe] axis. The top figures are for $\gamma = 1.5$ and $\gamma=2$ and have an unrealistic intrinsic 
scatter of 0.01 dex. The bottom figures are repeated but with an intrinsic scatter of 0.1 dex. The effects of 
clumping are clearly seen in both distributions.}
\end{figure}

\begin{figure}
\plotone{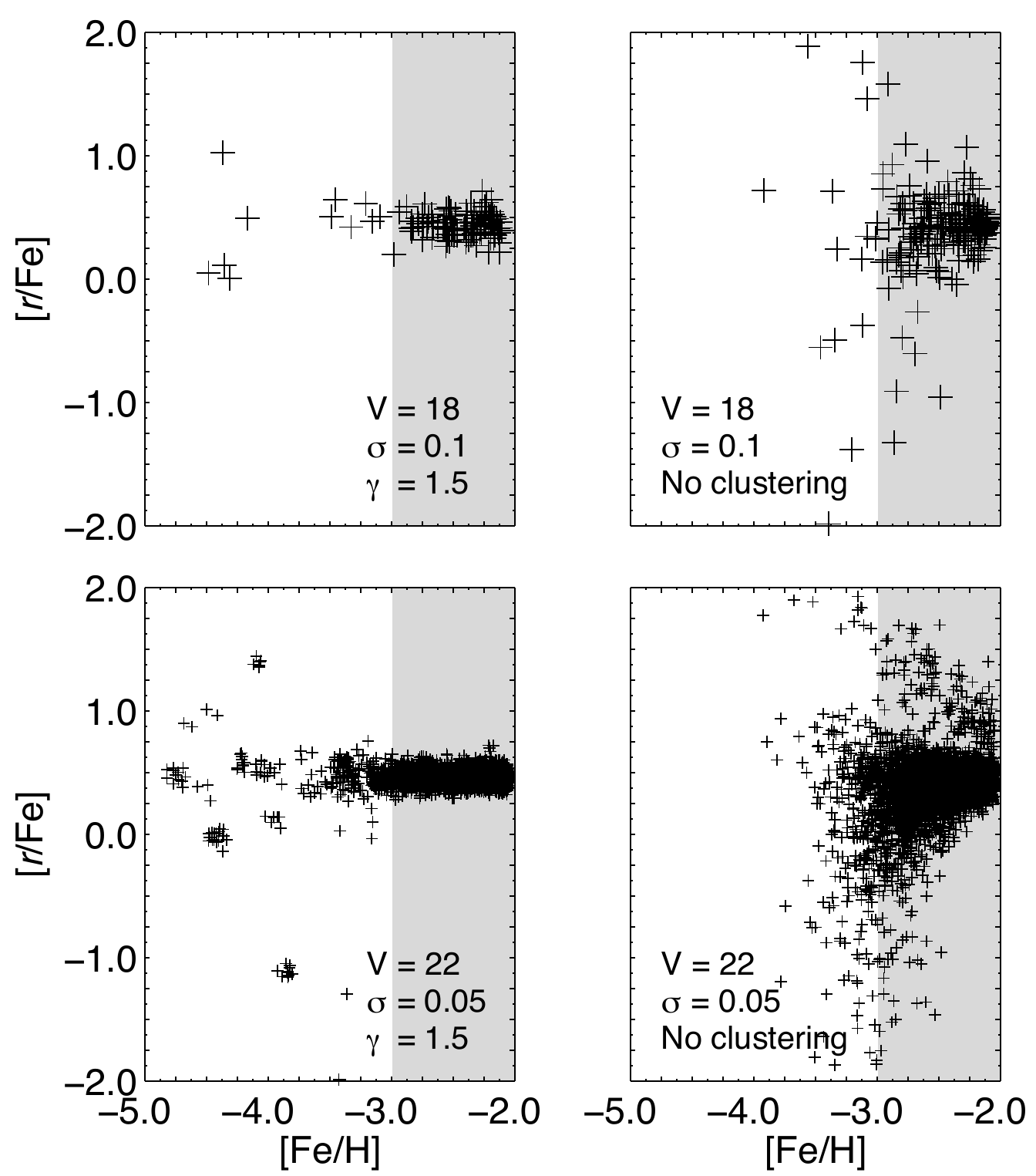}
\caption{ \label{noisy1}
A simulation of a targetted study of a nearby dwarf galaxy on an 8m (top) and 30m (bottom) telescope. 
The data points are drawn from Fig.~\ref{rFeH} for a galaxy with a stellar mass of $3\times 10^5$M$_\odot$
at a distance of 30 kpc (see Table~\ref{tab1}): (left) $\gamma=1.5$ (right) no clustering.
The simulated errors are 0.1 dex in the top figures and 0.05
dex in the bottom figures. There is evidence of clustering at [Fe/H] $<$ $-3.0$ from a sample of 10 stars
on an 8m class telescope; the clustering is easily detected in the 30m telescope experiment.}
\end{figure}

\begin{figure}
\plotone{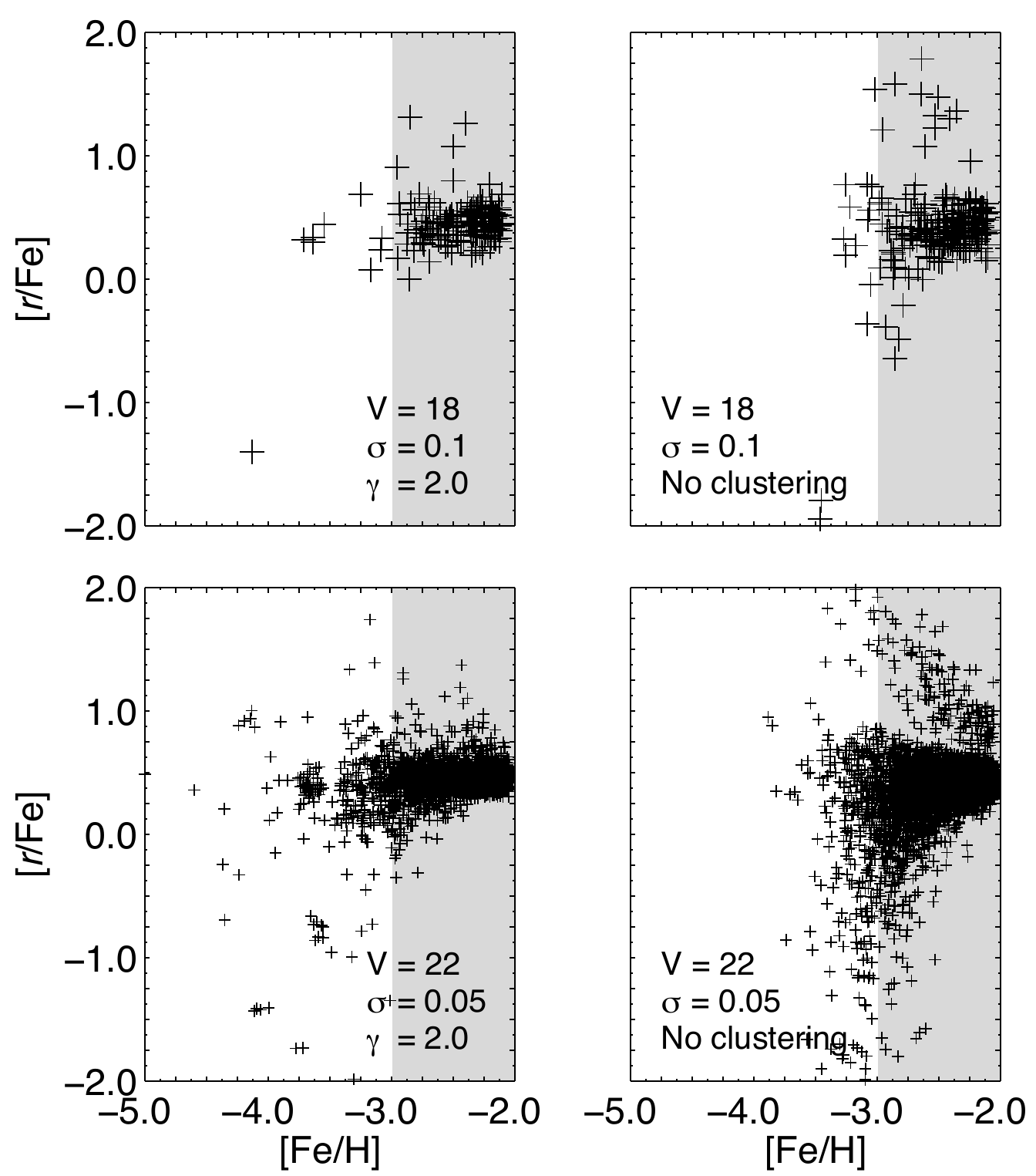}
\caption{ \label{noisy2}
A simulation of a targetted study of a nearby dwarf galaxy on an 8m (top) and 30m (bottom) telescope. 
The data points are drawn from Fig.~\ref{rFeH} for a galaxy with a stellar mass of $3\times 10^5$M$_\odot$
at a distance of 30 kpc (see Table~\ref{tab1}): (left) $\gamma=2.0$ (right) no clustering.
The simulated errors are 0.1 dex in the top figures and 0.05
dex in the bottom figures. There is evidence of clustering at [Fe/H] $<$ $-3.0$ from a sample of 10 stars
on an 8m class telescope; the clustering is easily detected in the 30m telescope experiment.}
\end{figure}

\appendix
\section{Group finding in an abundance space}

In the EnLink clustering scheme, each group is characterized by 
a maximum density $\rho_{\rm max}$ and a minimum density $\rho_{\rm min}$, and
these are used to define its significance $S$ as 
\begin{equation}
S=\frac{\ln(\rho_{\rm max})-\ln(\rho_{\rm  min})}{\sigma_{\ln(\rho)}}, 
\end{equation}
where $\sigma_{\ln(\rho)}$
is the standard deviation associated with the density
estimator and is a constant for a given dimensionality and $q_{den}$.

For Poisson-sampled data the distribution of density as estimated by
the code using the kernel scheme is log-normal and the variance satisfies
the relation
\begin{equation}
\sigma_{\ln(\rho)} = \sqrt{{\cal V}_d ||W||_2^2/q_{\rm den}} ,
\end{equation}
where $q_{den}$ is the number of neighbours employed for density
estimation, ${\cal V}_d$ the volume of a $d$-dimensional unit hypersphere and
$||W||_2^2$ the $L_2$ norm of the kernel function
\citep{sharma09}. For $q_{den}=6$ and $d=2$,  
$\sigma_{\ln(\rho)}$ evaluates to 0.4714.

Thus Enlink has two free parameters: (i) the significance threshold of the
group $S_{T}$, and (ii) the number of nearest neighbours $q_{den}$ used for
density estimation. The variable $q_{den}$ should be set to less than the minimum
desired size of the groups because the density is smoothed over a
scale of $q_{den}$ nearest neighbours such that the significance and hence the
probability of detecting groups of size less than $q_{den}$ is drastically reduced.
Additionally, $q_{link}={\rm min}(10,q_{den}-1)$ neighbours are used for
linking the groups
which means groups whose density peaks lie within  $q_{link}$ nearest neighbours
of each other cannot be separated.  
Since we are interested in identifying groups with less than 10 data
points, we set $q_{den}=6$; the case for $n_* \lesssim 6$ is discussed in
the next section.

\begin{figure}
  \centering \includegraphics[width=1.0\textwidth]{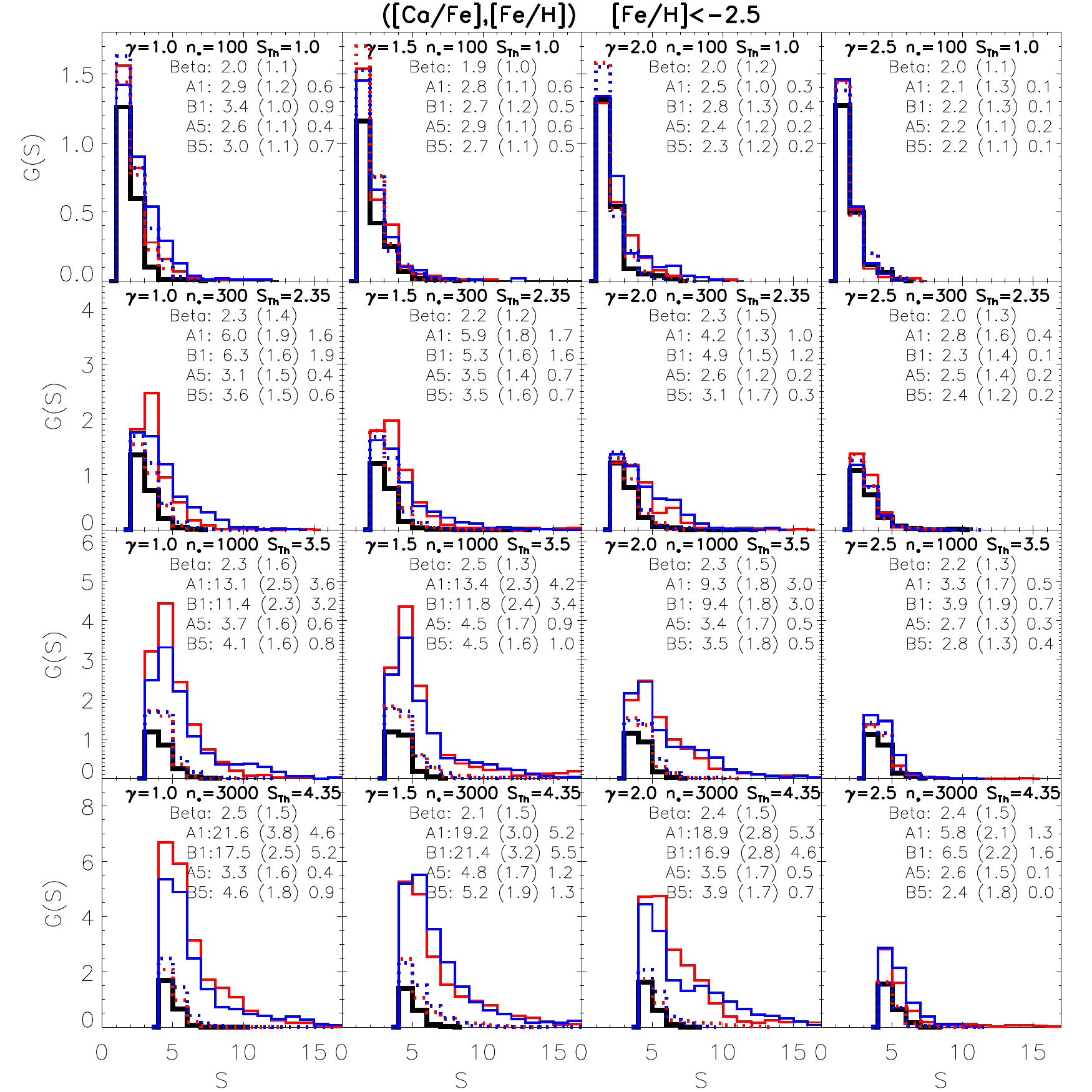}
\caption{ \label{cafe25}
The results of EnLink group finding applied to the models presented in Fig.~\ref{CaFeH} at a metallicity cut-off [Fe/H] $<$ $-2.5$. 
The models represent the 2D space ${\cal C}$([Fe/H], [Ca/Fe]) where the four rows show results for 
$n_* = 100, 300, 1000, 3000$ data points.  The four columns present results for $\gamma=1.0, 1.5, 2.0, 2.5$. 
In each panel, there are five distinct significance distribution functions $G(S)$ (equation 15); note that the vertical range increases
as $n_*$ increases (see \S 5.1).
The black histogram is the EnLink analysis of the control sample generated by the beta distribution in equation
\ref{eqn:beta} which, by definition, has no clustering. The solid blue and red histograms (A1, B1) are the analysis of the A and B realizations respectively
(0.01 dex uncertainty) shown in Fig.~\ref{CaFeH}; the dotted histograms (A5, B5) repeat the analysis for an abundance uncertainty of 0.05 dex. The statistical means and uncertainties (in brackets) for all distributions are 
given in the insets; the third value is the $C_S$ statistic (see text).}
\end{figure}

\begin{figure}
\centering \includegraphics[width=1.0\textwidth]{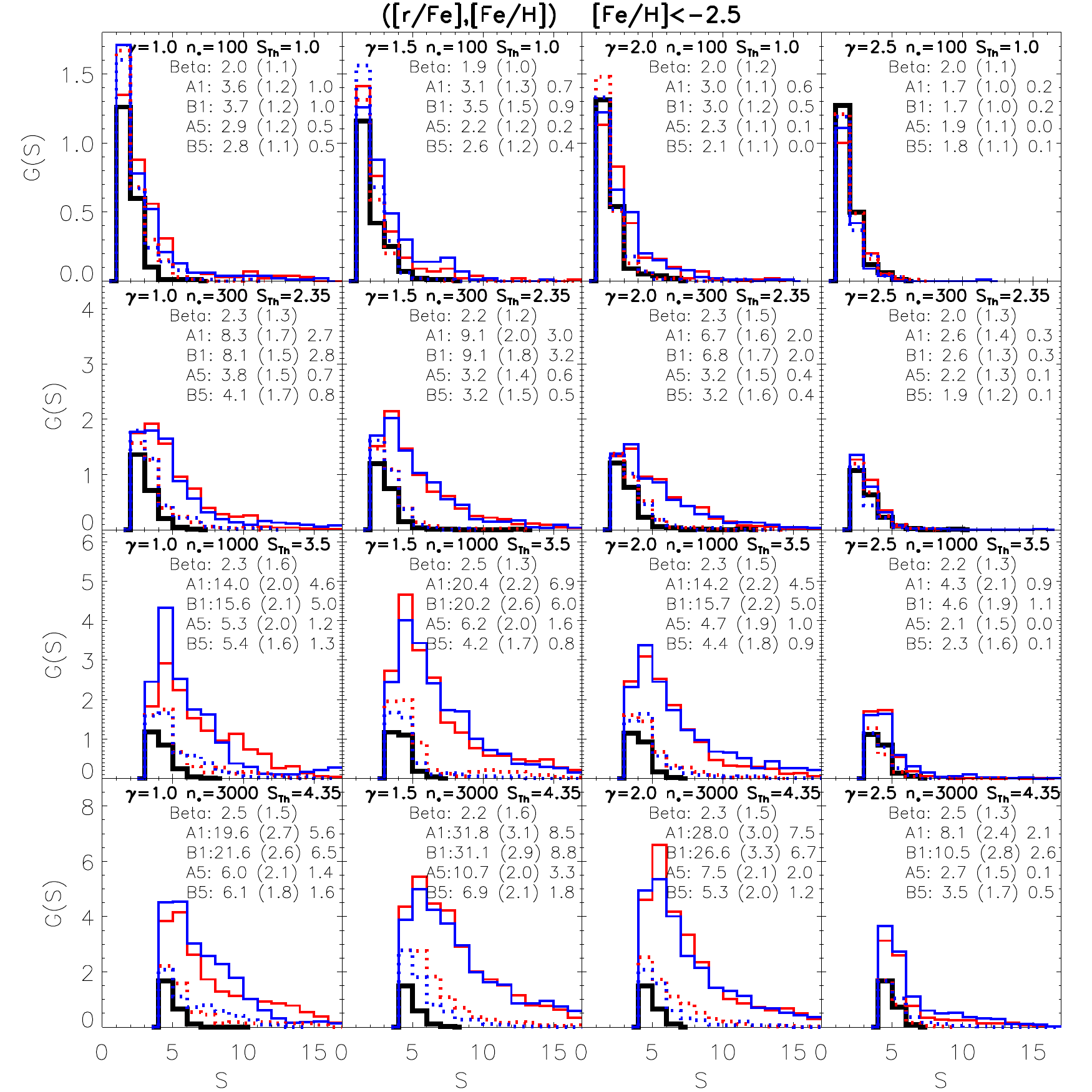}
\caption{ \label{rfe25}
The results of EnLink group finding applied to the models presented in Fig.~\ref{rFeH} at a metallicity cut-off [Fe/H] $<$ $-2.5$. 
The models represent the 2D space ${\cal C}$([Fe/H], [r/Fe]) where the four rows show results for 
$n_* = 100, 300, 1000, 3000$ data points.  The four columns present results for $\gamma=1.0, 1.5, 2.0, 2.5$. 
In each panel, there are five distinct significance distribution functions $G(S)$ (equation 15); note that the vertical range increases
as $n_*$ increases (see \S 5.1).
The black histogram is the EnLink analysis of the control sample generated by the beta distribution in equation
\ref{eqn:beta} which, by definition, has no clustering. The solid blue and red histograms (A1, B1) are the analysis of the A and B realizations respectively
(0.01 dex uncertainty) shown in Fig.~\ref{rFeH}; the dotted histograms (A5, B5) repeat the analysis for an abundance uncertainty of 0.05 dex. The statistical means and uncertainties (in brackets) for all distributions are 
given in the insets; the third value is the $C_S$ statistic (see text).}
\end{figure}

Our initial analysis is for data points with [Fe/H] $<$ -2.5. This includes a
crowded, high-density region of the abundance plane which skews the
clustering analysis considerably. However, this may be an important
regime if the first chemical elements arise from pair instability supernovae
(Karlsson et al 2008). The significance of spurious
groups is distributed almost as a Gaussian distribution and hence 
the significance threshold $S_{T}$ is used to suppress these groups. 
An improved version of the empirical formula \citep{sharma09} for the expected number 
of spurious groups, valid for
$S_{T}>1$, is given by 
\begin{equation}
G(S>S_T)=\left(1-{\rm erf}\left(\frac{S_T f_{dq}}{\sqrt{2}}\right)\right)\frac{0.4n_*}{q_{\rm den}}
\label{eqn:gs}
\end{equation}
where $f_{dq}=0.5 \sqrt{d(1-2.3/q_{\rm den})}$ is a small correction
term and $n_*$ is the number of data points.\footnote{Since the distribution in $S$ for spurious 
groups is roughly normal (equation 15), $S$ can be loosely interpreted as the statistical z-score 
for finding a group per unit data point.} In Figs.~\ref{cafe25} and \ref{rfe25}, 
we plot the significance distribution $S$ of
identified groups for data sets with different values of $n_*$ and $\gamma$ 
as labelled on the plots. 

We set $S_T$ such that the expected number of spurious groups
(false positives)
with $S>S_{T}$ is about 2 in all data sets which we calibrate from the
beta models. $S_{T}$ is set to 1.0, 2.35, 3.5
and 4.35 for data with $n_*$ as 100, 300, 1000 and 3000 respectively.
This number is increasing because, for a fixed $q_{den}$ and data dimensionality,
the number of spurious groups due to Poisson noise increases linearly
with the total number of data points $n_*$.

Each panel shows the outcome of five EnLink analyses: the cluster-less 
beta model (solid black line); the simulated models A1 and B1 having unrealistic
abundance errors 0.01 dex shown as solid red and blue curves respectively;
the simulated models A5 and B5 having abundance errors 0.05 dex shown
as dotted red and blue lines respectively. In each panel, the mean number 
of identified groups $G$ along with its dispersion $\sigma_{G}$ 
is also given. The mean and dispersion were calculated using 
100 random realizations of each data type. The plotted distributions
are also averaged over these 100 random realizations.
We also provide a measure of the
statistical significance of detecting clusters in a data set
as follows. If $G_{B}$ is the number of groups as predicted by the 
smooth beta model, the statistical significance of clustering in 
a model is given by 
\begin{equation}
C_S=\frac{\langle G\rangle - \langle G_B\rangle}{\sqrt{\sigma_{G_B}^2+\sigma_{G}^2}} 
\end{equation}
The third column shows the value of $C_S$ for each data set.

It can be seen in Fig.~\ref{cafe25} and \ref{rfe25} that the expected 
number of spurious groups for the beta models are nearly independent of the value of 
$\gamma$, which is a consequence of the adaptive metric scheme.
Specifically, EnLink uses the concept of a locally adaptive metric, which is used for 
calculating densities and nearest neighbours of data points
(a refinement over earlier developments by Ascasibar \& Binney [2005])
to increase the efficiency of detecting clusters in multi-dimensional spaces.
If instead clustering is performed using a Euclidean metric on raw data, 
the significance distribution of spurious groups due to Poisson noise 
is found to vary with the distribution of points in the
abundance space, e.g., the beta models with different values of $\gamma$.

In the limit of few data points, the locally adaptive metric
scheme is equivalent to using a metric which is given by the inverse 
of the dispersion along each dimension. As a check we also performed 
the analysis using this simpler scheme and found equivalent 
results, thereby demonstrating the robustness of our group-finding
analysis. For the sake of accuracy, we use the beta models to evaluate
the significance of clustering, but strictly speaking this is not 
required and any Poisson sampled cluster-less data would also suffice. 
In fact, the empirically derived formula given by
Equation~(\ref{eqn:gs}) can also be used directly to predict 
the number of spurious groups.

Next we compare the parameter $C_S$ for the different cases. 
It can be seen that as $n_*$ and $\gamma$ increase, $C_S$ increases also. 
First we look at cases with unrealistic measurement errors of 0.01 dex. 
It is clear that the $\gamma=2.5$ models are nearly undetectable for all values of 
$n_*$ and closely resemble the beta models. For $\gamma=2.0$, a minimum
value of $n_*=300$ is needed for $C_S \gtrsim 1$. 
For lower values of $\gamma$, signatures of clustering are 
visible with as few as 100 points. Increasing the abundance errors 
to 0.05 dex significantly affects our ability
to detect clusters. When sufficient number of data points are present,
the A5 and B5 models perform better for $\gamma=1.5$ and $2.0$ 
as compared to $\gamma=1$. This is because the clusters 
in these cases are more in number, are spread over a larger area, 
are less crowded, and hence easier to detect.
   
\begin{figure}
  \centering \includegraphics[width=1.0\textwidth]{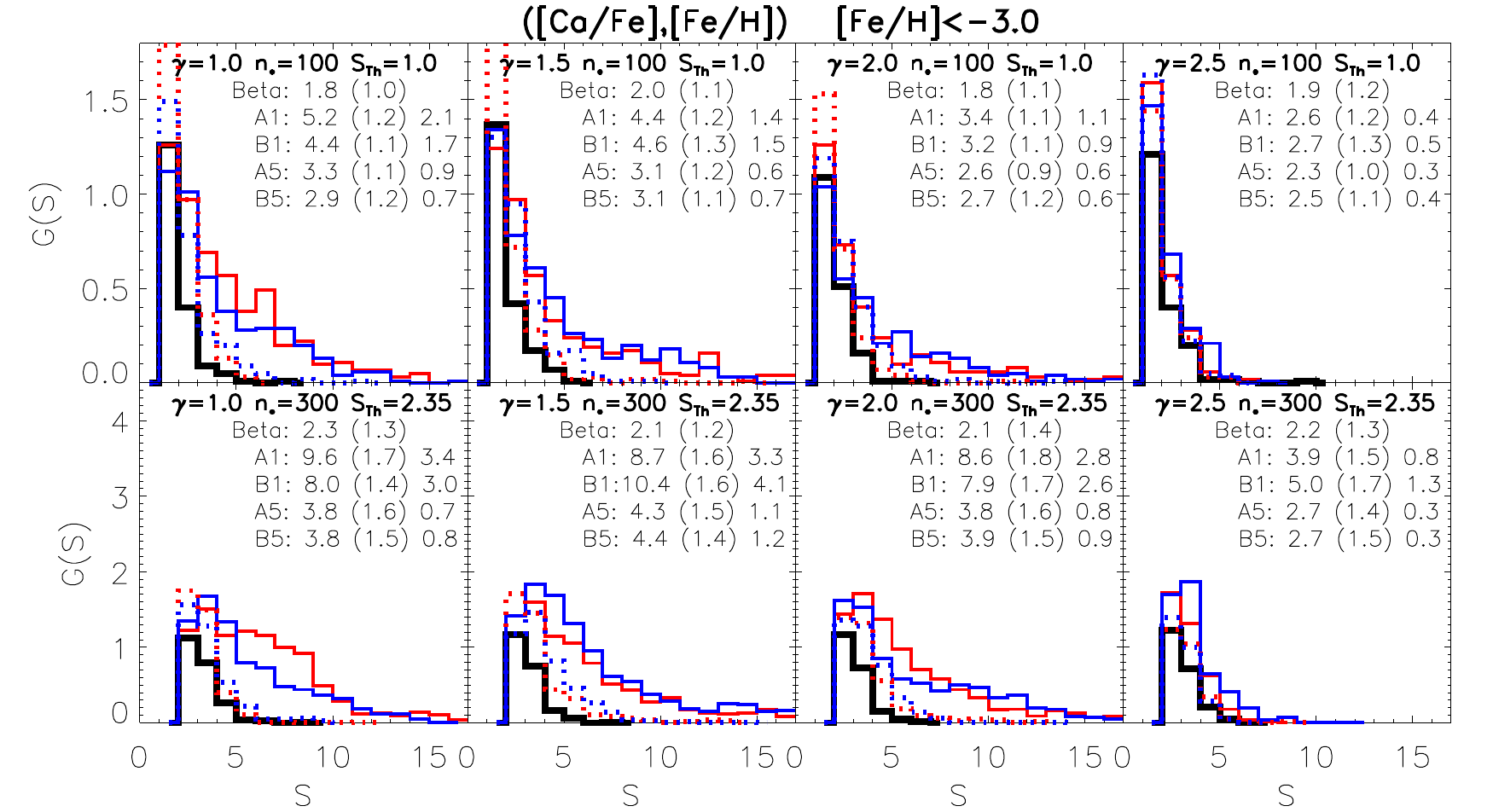}
\caption{ \label{cafe30}
The results of EnLink group finding applied to the models presented in Fig.~\ref{CaFeH} for
[Fe/H] $<$ $-3.0$. The two rows show results for $n_*$ $=$ 100 and 300 data points. The four columns
present results for $\gamma=1.0, 1.5, 2.0, 2.5$. The histograms are defined in Fig.~\ref{cafe25}. The
effects of clustering are now more apparent, even in the limit of small statistics.}
\end{figure}

\begin{figure}
\centering \includegraphics[width=1.0\textwidth]{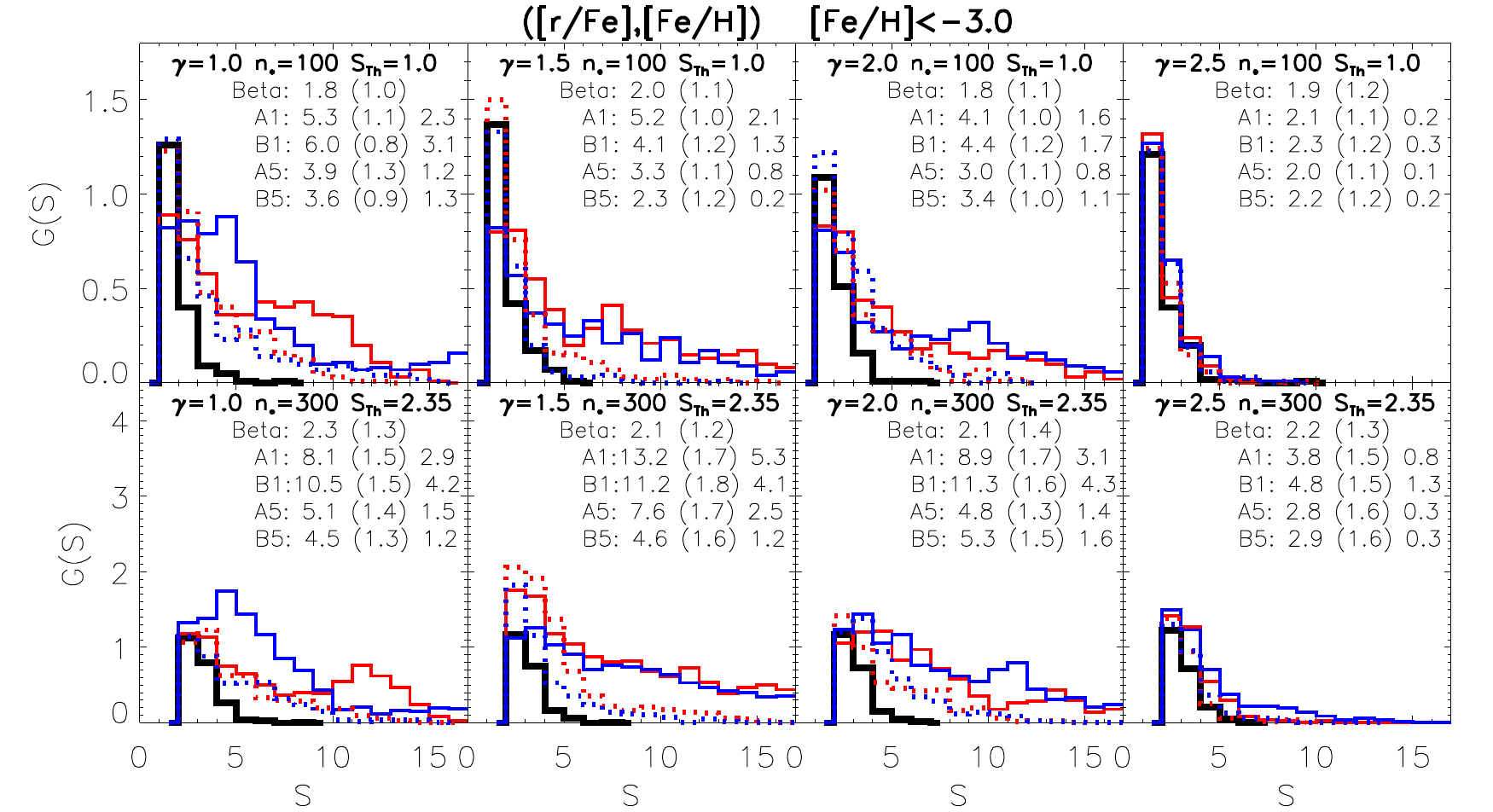}
\caption{ \label{rfe30}
The results of EnLink group finding applied to the models presented in Fig.~\ref{rFeH} for
[Fe/H]$<-3.0$. The two rows show results for $n_*$ $=$ 100 and 300 data points. The four columns
present results for $\gamma=1.0, 1.5, 2.0, 2.5$. The histograms are defined in Fig.~\ref{rfe25}. The
effects of clustering are now more apparent, even in the limit of small statistics.}
\end{figure}

\begin{figure}
\centering \includegraphics[width=1.0\textwidth]{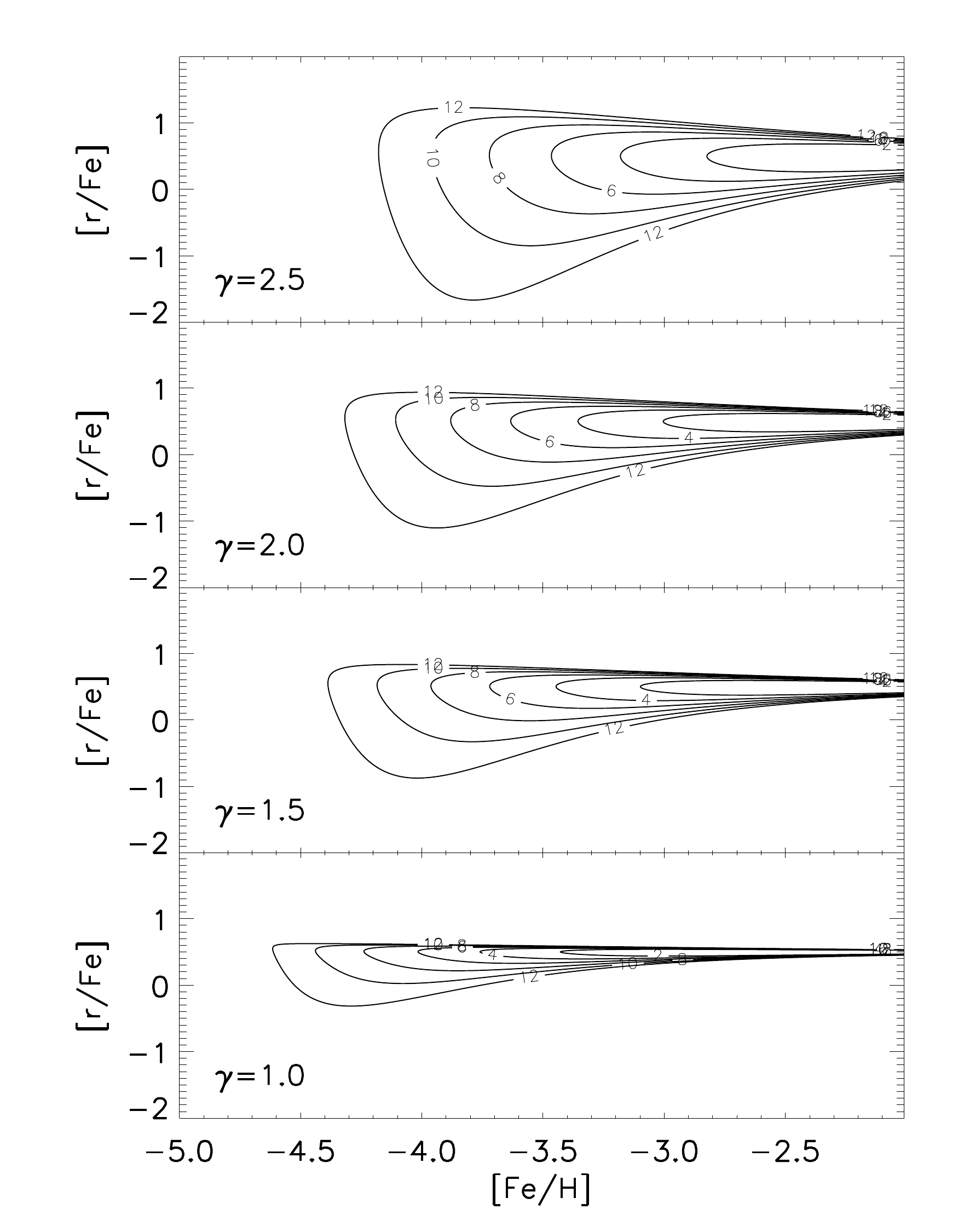}
\caption{ \label{Sdist}
The statistical significance distribution $S$ (equation 13) for the four clustered models presented in Fig.~\ref{rFeH}.
The contours are (from right to left): $S=2,4,6,8,10,12$. Note, for example, groupings of data points are more likely
near the mean value of [r/Fe] at low [Fe/H] for low $\gamma$.
Clustered abundance data points away from the mean are favoured by higher values of $\gamma$.
}
\end{figure}

Since clustering is most prominent for lower values of [Fe/H], we
also investigated data sets with a lower metallicity cut-off, [Fe/H] $<$ -3.0. These
results are shown in Figs.~\ref{cafe30} and \ref{rfe30}.
The parameter $C_S$ is found to increase in general for all cases, unlike
what is seen in the [Fe/H] $<$ -2.5 analysis (Figs.~\ref{cafe25} and \ref{rfe25}).
It is striking how much better EnLink
performs on these simulations which is to be expected given the broader intrinsic
dispersions of [r/Fe] in the models. In light of these results, another way to look
at Fig.~\ref{rFeH} is the significance $S$ of clustering over the abundance
plane. In Fig.~\ref{Sdist}, we show the $S$ contours for the four different values
of $\gamma$ with lower significance regions to high [Fe/H]. Clustered abundance 
data points near the mean value of [r/Fe] are more likely at low [Fe/H] for low values of $\gamma$. 
Clustered abundance data points away from the mean are favoured by higher values of $\gamma$.

\begin{figure}
\centering \includegraphics[width=1.0\textwidth]{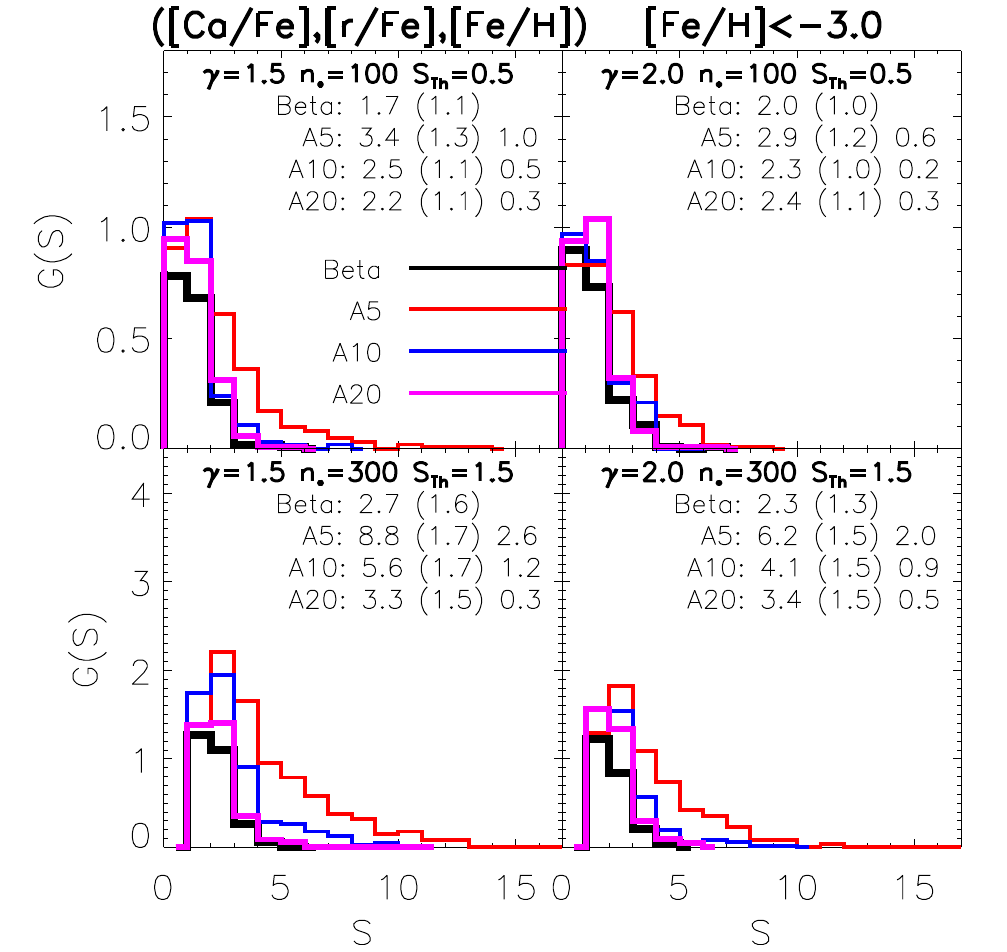}
\caption{ \label{rcafe30}
The results of EnLink group finding applied to the 3D space ${\cal C}$([Fe/H], [Ca/Fe], [r/Fe])
for [Fe/H] $<$ -3.0. The two rows show results for N $=$ 100 and 300 data points. The two columns
present results for $\gamma=1.5, 2.0$. The histograms are defined in Fig.~\ref{rfe25}. 
The simulated measurement errors are now 0.05 (red), 0.10 (blue) and 0.20 (purple) dex. The
effects of clustering are apparent, even in the limit of small statistics and with larger 
measurement errors.}
\end{figure}

\end{document}